\documentclass[11pt,oneside,letterpaper]{article}

\usepackage{graphicx}
\usepackage{grffile}

\usepackage{color} 
\usepackage{parskip}
\usepackage{float}

\usepackage{xr}
\usepackage{geometry}
\usepackage[labelfont=bf,labelsep=period,font=small]{caption}

\usepackage{natbib}

\usepackage{amsmath}

\usepackage{authblk}

\renewcommand\Affilfont{\small \normalfont}
\makeatletter
\renewcommand\AB@affilsepx{, \protect\Affilfont}
\makeatother

\usepackage{ulem}
\definecolor{purple}{rgb}{0.459,0.109,0.538}

\newcommand{\chiSq}{\chi^{2}_{df}}
\newcommand{\dtmrca}{\Delta_\mathrm{TMRCA}}
\newcommand{\undtmrca}{\delta_\mathrm{TMRCA}}
\newcommand{\dspr}{d_\mathrm{SPR}}

\title{\vspace{1.0cm} \LARGE \bf Reassortment between influenza B lineages and the emergence of a co-adapted PB1-PB2-HA gene complex}

\author[1]{Gytis Dudas}
\author[2]{Trevor Bedford}
\author[1,3]{Samantha Lycett}
\author[1,4,5]{Andrew Rambaut}

\affil[1]{Institute of Evolutionary Biology, University of Edinburgh, Edinburgh, UK}
\affil[2]{Vaccine and Infectious Disease Division, Fred Hutchinson Cancer Research Center, Seattle, WA, USA}
\affil[3]{Institute of Biodiversity Animal Health and Comparative Medicine, University of Glasgow, Glasgow, UK}
\affil[4]{Fogarty International Center, National Institutes of Health, Bethesda, MD, USA}
\affil[5]{Centre for Immunology, Infection and Evolution at the University of Edinburgh, Edinburgh, UK}

\date{\today}

\begin{document}
\maketitle

\begin{abstract}

Influenza B viruses make a considerable contribution to morbidity attributed to seasonal influenza. 
Currently circulating influenza B isolates are known to belong to two antigenically distinct lineages referred to as B/Victoria and B/Yamagata. 
Frequent exchange of genomic segments of these two lineages has been noted in the past, but the observed patterns of reassortment have not been formalized in detail.
We investigate inter-lineage reassortments by comparing phylogenetic trees across genomic segments.
Our analyses indicate that of the 8 segments of influenza B viruses only PB1, PB2 and HA segments maintained separate Victoria and Yamagata lineages and that currently circulating strains possess PB1, PB2 and HA segments derived entirely from one or the other lineage; other segments have repeatedly reassorted between lineages thereby reducing genetic diversity.
We argue that this difference between segments is due to selection against reassortant viruses with mixed lineage PB1, PB2 and HA segments.
Given sufficient time and continued recruitment to the reassortment-isolated PB1-PB2-HA gene complex, we expect influenza B viruses to eventually undergo sympatric speciation.

\end{abstract}

\pagebreak

\section*{Introduction}
Seasonal influenza causes between 250,000 and 500,000 deaths annually and comprises lineages from three virus types (A, B and C) co-circulating in humans, of which influenza A is considered to cause the majority of seasonal morbidity and mortality \citep{flufactsheet}. Occasionally influenza B viruses become the predominant circulating virus in some locations, for example in the 2012/2013 European season as many as 53\% of influenza sentinel surveillance samples tested positive for influenza B \citep{ECDC1213}. 

Like other members of \textit{Orthomyxoviridae}, influenza B viruses have segmented genomes, which allow viruses co-infecting the same cell to exchange segments, a process known as reassortment. 
Influenza A viruses are widely considered to be a major threat to human health worldwide due to their ability to cause pandemics in humans via reassortment of circulating human strains with non-human influenza A strains. 
Although influenza B viruses have been observed to infect seals \citep{osterhaus2000,bodewes2013} through a reverse zoonosis, they are thought to primarily infect humans and are thus unlikely to exhibit pandemics due to the absence of an animal reservoir from which to acquire antigenic novelty. 
Both influenza A and B evolve antigenically through time in a process known as antigenic drift, in which mutations to the haemagglutinin (HA) protein allow viruses to escape existing human immunity and persist in the human population, leading to recurrent seasonal epidemics \citep{burnet1955,hay2001,bedford2014}.

Currently circulating influenza B viruses comprise two distinct lineages -- Victoria and Yamagata (referred to as Vic and Yam, respectively) -- named after strains B/Victoria/2/87 and B/Yamagata/16/88, that are thought to have genetically diverged in HA around 1983 \citep{rota1990}. 
These two lineages now possess antigenically distinct HA surface glycoproteins \citep{kanegae1990,rota1990,nerome1998,nakagawa2002,ansaldi2003} allowing them to co-circulate in the human population.
Phylogenetic analysis of evolutionary rate, selective pressures and reassortment history of influenza B has shown extensive and often complicated patterns of reassortment between all segments of influenza B viruses both between and within the Vic and Yam lineages \citep{chen2008}.

\begin{figure}[h]
 \centering		
	\includegraphics[width=0.45\textwidth]{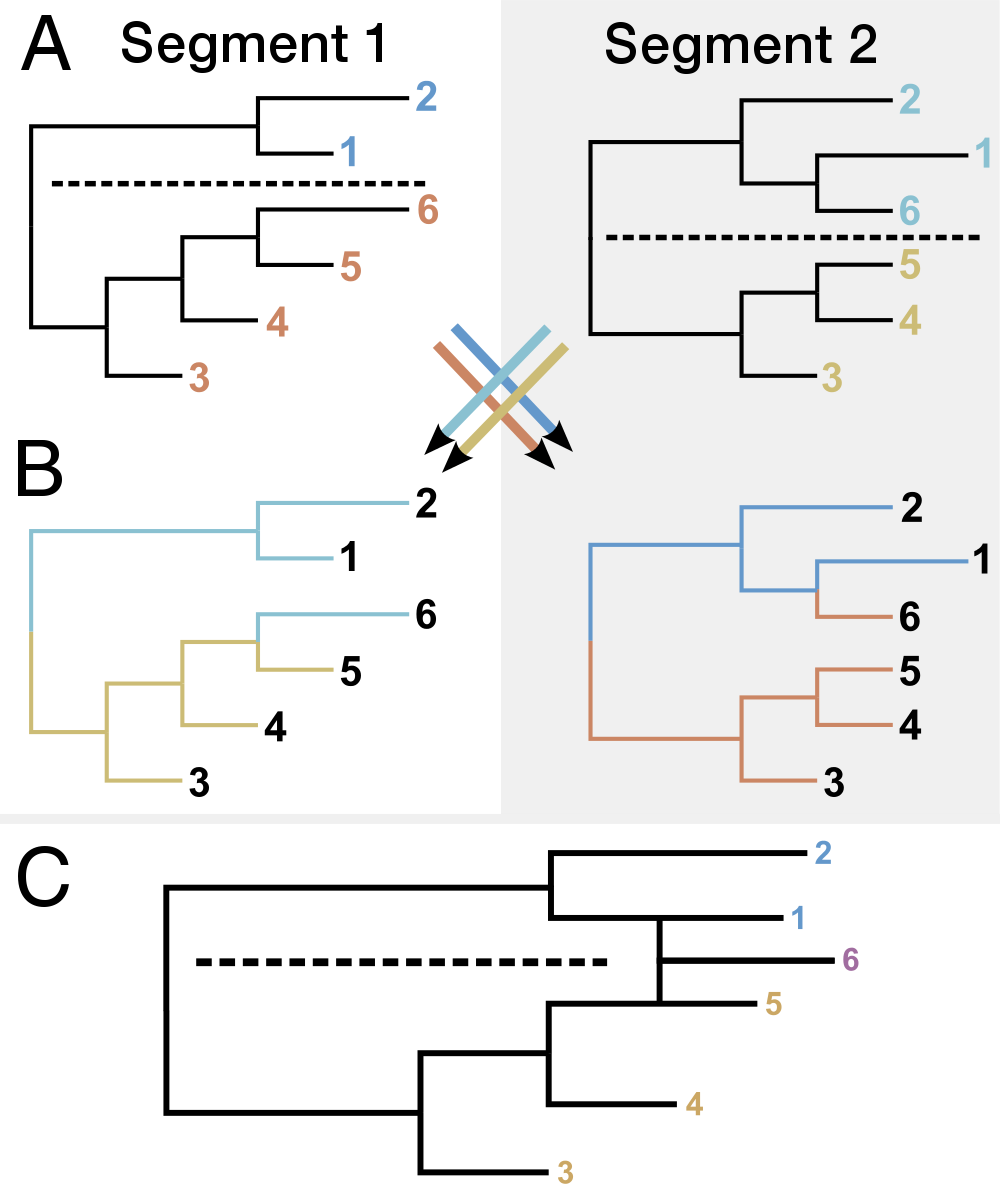}
	\caption{\textbf{Schematic analysis of reassortment patterns.}
	A) We begin by assigning sequences falling on either side of a specified bifurcation within each segment tree to different lineages, in this case, the Victoria and Yamagata bifurcation that occurred in the early 1980s.
	B) We then transfer lineage labels from one tree to the same tips in another tree.
	Transitions between labels along this second tree thus indicate reassortment events that combine lineages falling on different sides of the Vic/Yam bifurcation in the first tree.
	C) A reassortment graph depiction shows that tip number 6 is determined to be a reassortant based on B).
	}
	\label{methodFig}
\end{figure}

Here, we extend previous methods to reveal an intriguing pattern of reassortment in influenza B.
In our approach, membership to either the Victoria or Yamagata lineage in one segment is used to label the individual isolates in the tree of the other segments.
By modelling the transition between labels on a phylogenetic tree, reassortment events which result in the replacement of one segment's lineage by another show up as label changes along a branch (Figure \ref{methodFig}).
We use this method to reconstruct major reassortment events and quantify reassortment dynamics over time in a dataset of 452 influenza B genomes, and conduct secondary analyses in a dataset of 1603 influenza B genomes.

We show that despite extensive reassortment, three of the eight segments -- two segments coding for components of the influenza B virus polymerase, PB1 and PB2, and the surface glycoprotein HA -- still survive as distinct Victoria and Yamagata lineages, which appear to be co-dependent to the point where virions which do not contain PB1, PB2 or HA segments derived entirely from either the Vic or the Yam lineage have rarely been isolated and only circulate as transient lineages once isolated.
In other segments (PA, NP, NA, MP and NS) a single lineage has introgressed into the opposing background and replaced the previous lineage: all currently circulating influenza B viruses have PA, NP, NA and MP segments derived from Yamagata lineage and NS segments derived from Victoria lineage.

\section*{Results}

\subsection*{Analysis of reassortment patterns across Victoria and Yamagata lineages}
\begin{figure}[h]
\centering
\includegraphics[width=0.95\textwidth]{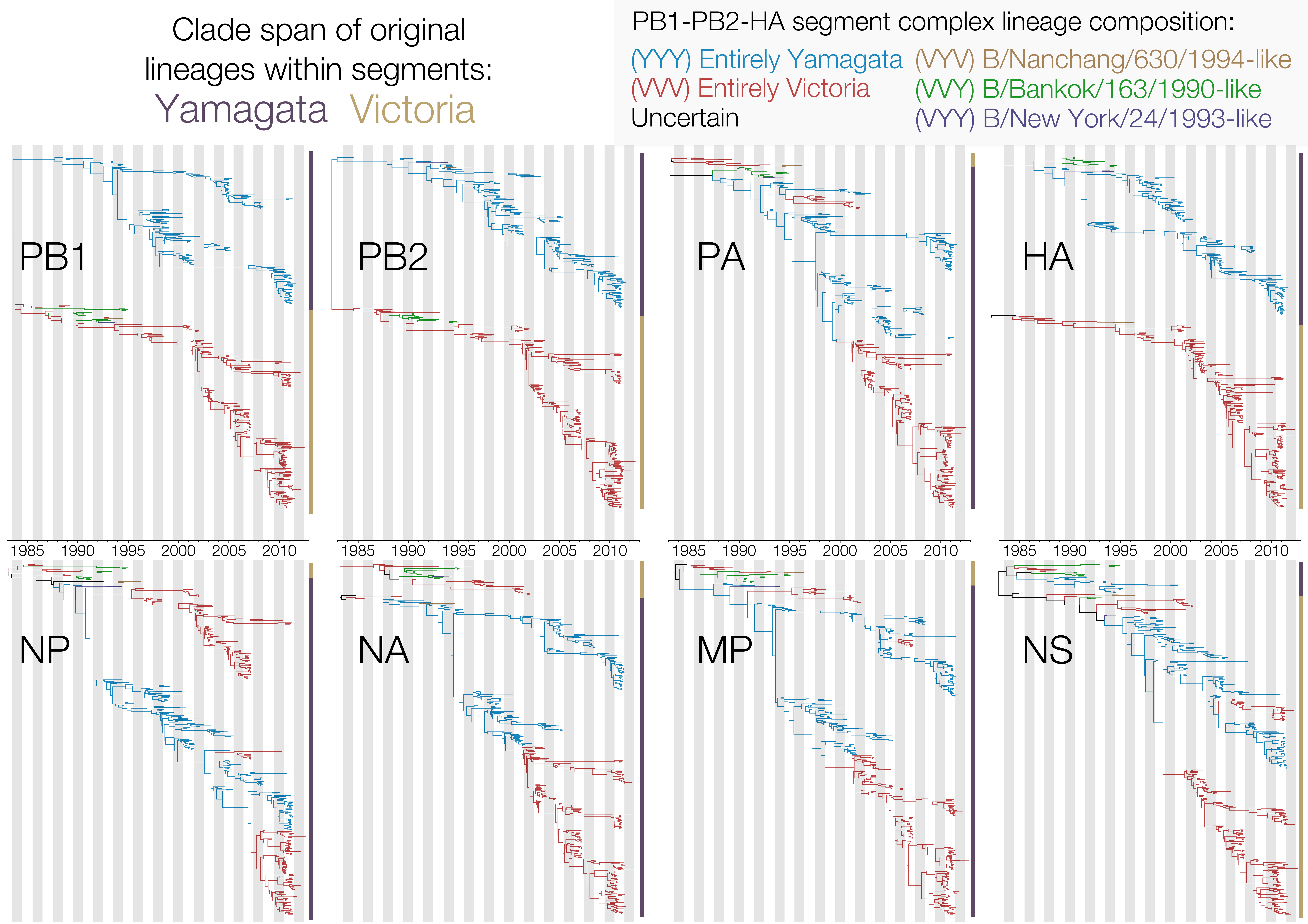}
\caption{\textbf{Maximum clade credibility (MCC) trees of all 8 genome segments of influenza B viruses isolated since 1980.}
Trees are coloured based on inferred PB1-PB2-HA lineage.
Vertical bars indicate the original Victoria and Yamagata lineages within each segment.
Each tree is the summarised output of a single analysis comprised of 9000 trees sampled from the posterior distribution of trees.}
\label{genomeGrid}
\end{figure}

\begin{figure}[h]
	\centering		
	\includegraphics[width=0.75\textwidth]{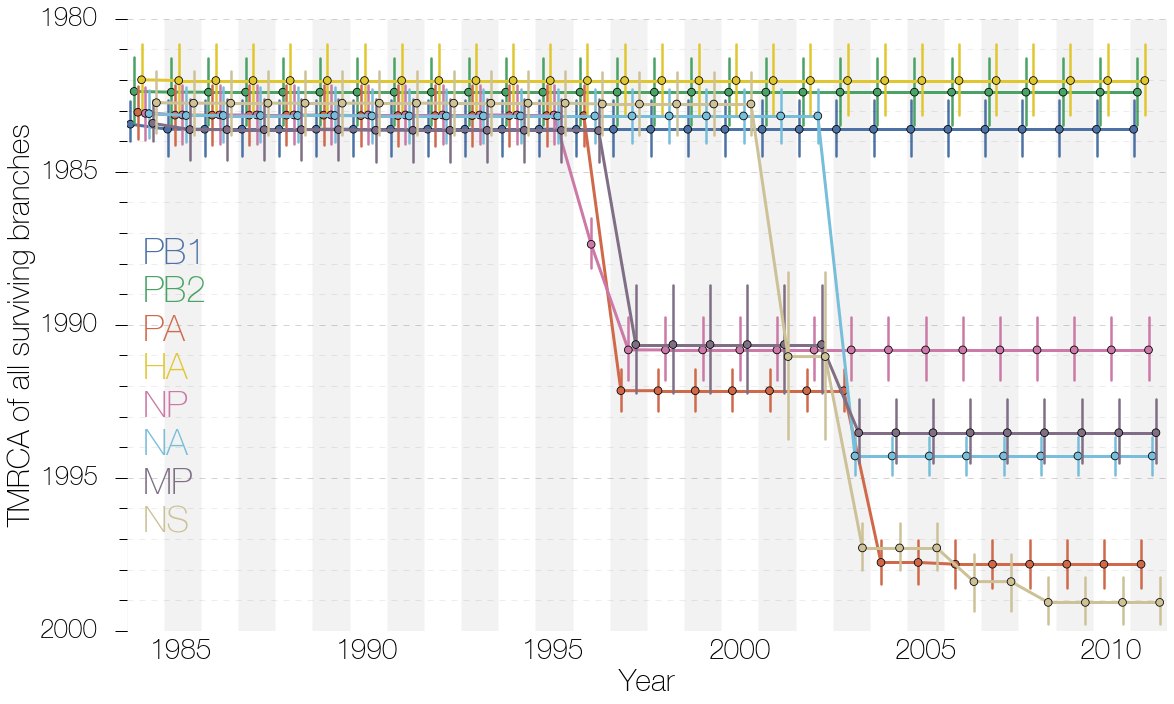}
	\caption{\textbf{Oldest TMRCA of all surviving branches over time.}
PA, NP, NA, MP and NS segments of influenza B viruses show periodic increases in TMRCAs of all surviving branches (indicative of diversity loss), suggesting lineage turnover.
PB1, PB2 and HA segments, on the other hand, maintain the diversity dating back to the initial split of Vic and Yam lineages.
Each point is the mean time of most recent common ancestor (TMRCA) of all surviving lineages existing at each time slice through the tree and vertical lines indicating uncertainty are 95\% highest posterior densities (HPDs).}
	\label{tmrcaOT}
\end{figure}

The differentiation into Vic and Yam lineages can be seen in all segments (Figure \ref{genomeGrid}) and is followed by inter-lineage reassortment events.
In the phylogenetic trees of the PA, NP, NA, MP and NS segments either the Victoria or Yamagata lineage has become the `trunk' of the tree, with present-day viruses deriving entirely from the Victoria or Yamagata lineage (yellow vs purple bars in Figure \ref{genomeGrid}) following reassortment.
However, the Victoria and Yamagata lineages of PB1, PB2 and HA segments continue to co-circulate to this day.
Periodic loss of diversity in PA, NP, NA, MP and NS segments is consistent with introgression of one lineage into the other in those segments, while maintenance of parallel Victoria and Yamagata lineages results in continually increasing diversity in segments PB1, PB2 and HA (Figure \ref{tmrcaOT}).
The PB1, PB2 and HA segments from present-day viruses maintain a common ancestor in $\sim$1983 and thus accumulate genetic diversity since the split of those segments into Vic and Yam lineages, while other segments often lose diversity with ancestors to present-day viruses appearing between $\sim$1991 and $\sim$1999.

\begin{figure}[h]
	\centering		
	\includegraphics[width=0.75\textwidth]{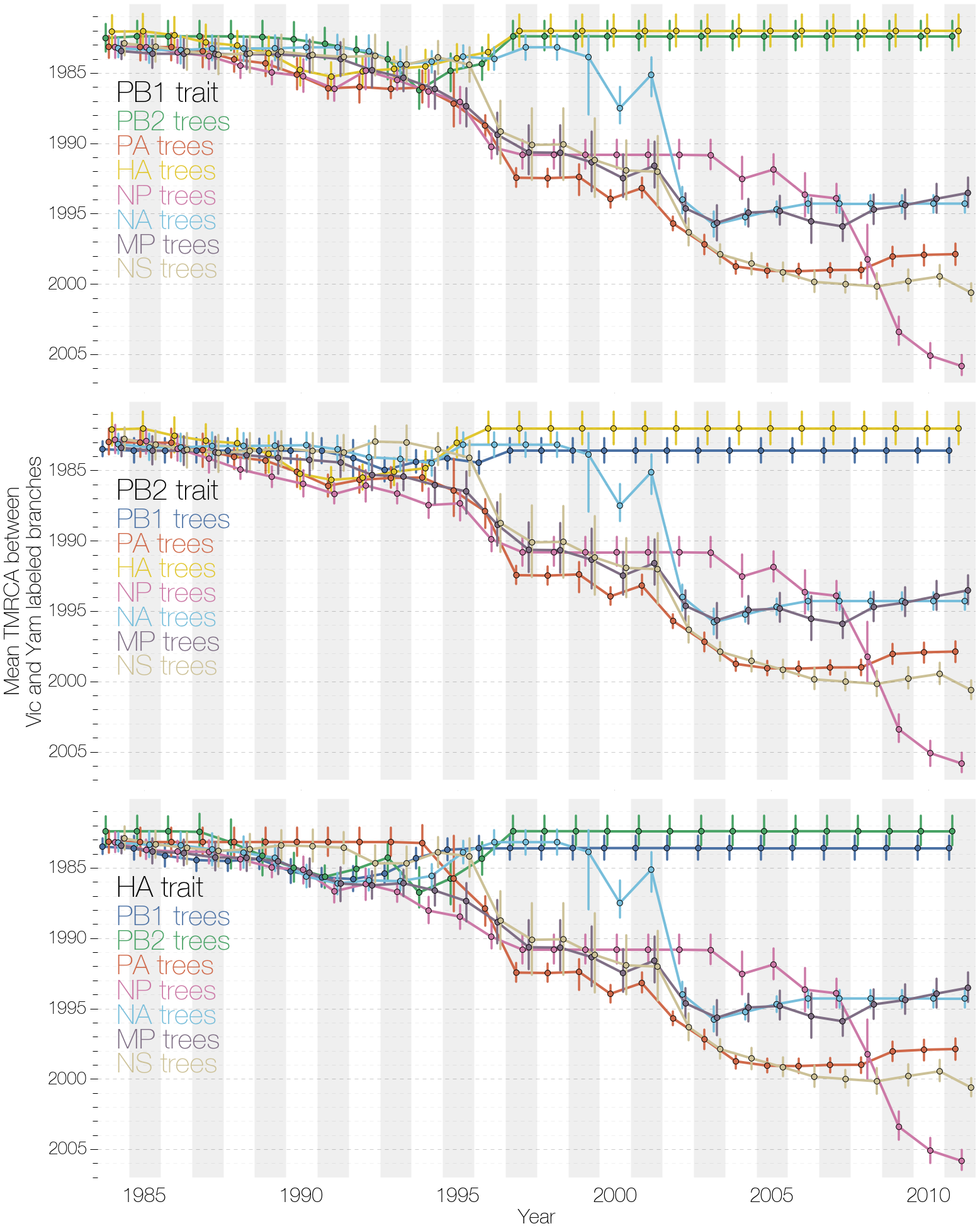}
	\caption{\textbf{Mean pairwise TMRCA between Vic and Yam branches under PB1, PB2 and HA label sets.}
PB1, PB2 and HA segment labels indicate that these segments show reciprocal preservation of diversity, which dates back to the split of Vic and Yam lineages.
All other segments show increasingly more recent TMRCAs between branches labelled as Vic and Yam in PB1, PB2 and HA label sets.
All vertical lines indicating uncertainty are 95\% highest posterior densities (HPDs).}
	\label{betweenDiversity}
\end{figure}

By measuring mean pairwise diversity between branches in each tree that were assigned either a Vic or Yam label in other segments, we look for reductions in between-lineage diversity, which indicate that an inter-lineage reassortment event has taken place (Figure \ref{betweenDiversity}).
This method gives a quantitative measure of reassortment-induced loss of diversity between Victoria and Yamagata lineages in two trees, although care should be taken when interpreting the statistic, as it does not correspond to any real TMRCAs in the tree, but can be interpreted as mean coalescence date between Vic and Yam lineages of PB1, PB2 and HA segments in all other trees.
We focus only on PB1, PB2 and HA lineage labels, since all other segments eventually become completely derived from either the Vic or the Yam lineage.
Losses of diversity (represented by more recent mean pairwise TMRCAs between Vic and Yam labels) in Figure \ref{betweenDiversity} indicate that every segment has reassorted with respect to the Victoria and Yamagata lineages of PB1, PB2 and HA segments.
However, we also see that the labels for these 3 segments show reciprocal preservation of diversity after 1997.
This suggests that after 1997 no reassortment events have taken place between Victoria and Yamagata lineages of PB1, PB2 and HA segments and their lineage labels only `meet' at the root.
We do see reduced diversity between Vic and Yam labels of PB1, PB2 and HA segments in a time period close to the initial split of Vic and Yam lineages (1986--1996).
These reductions in diversity represent small clades with reassortant PB1-PB2-HA constellations, which go extinct by 1997 (see Figure \ref{genomeGrid}).
We also observe that the assignment of these 3 segment labels to branches of other segment trees is very similar and often identical after 1997.
This suggests that PB1, PB2 and HA lineage labels switch simultaneously in all trees after 1997.

\begin{figure}[h]
	\centering	
	\includegraphics[width=0.65\textwidth]	{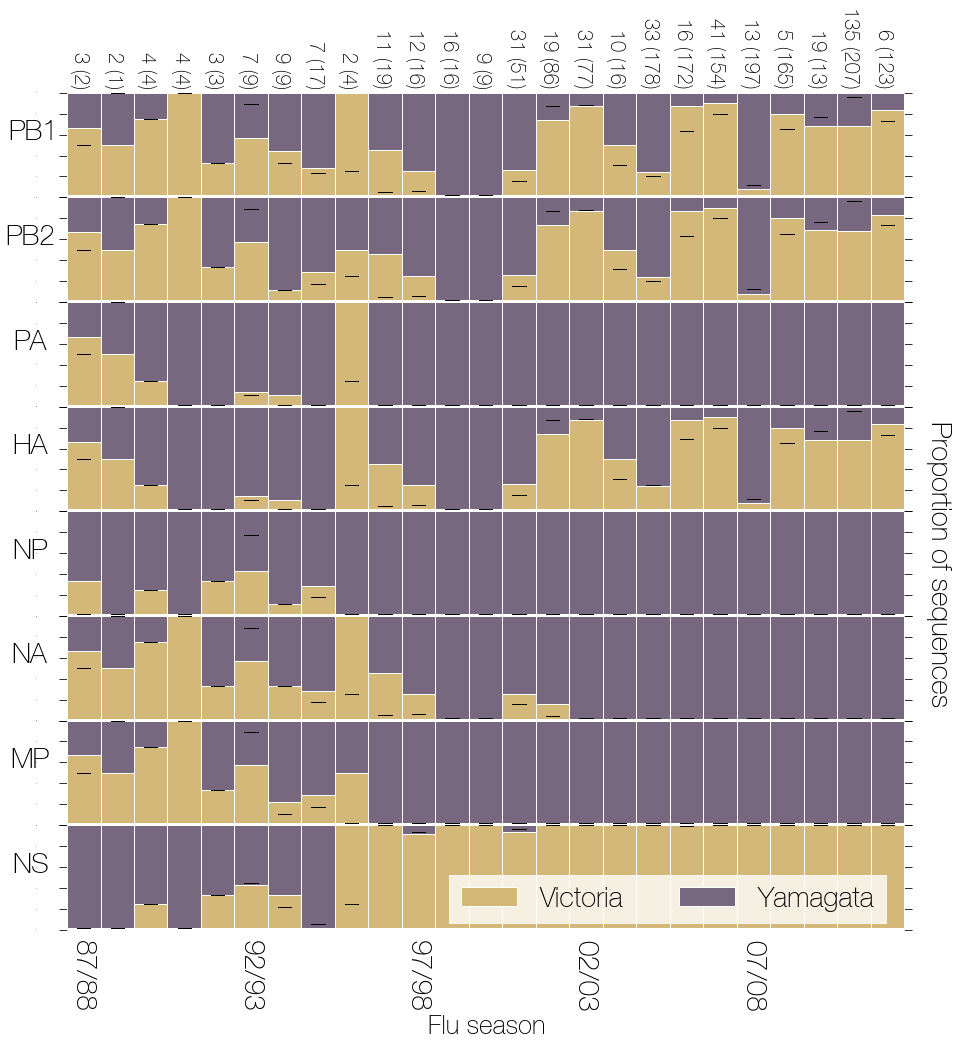}
	\caption{\textbf{Ratio of Vic and Yam sequences in the dataset.}
The ratio of Victoria (yellow) to Yamagata (purple) sequences in each segment from the primary dataset over time.
Black lines indicate where this ratio lies in the larger secondary dataset.
Numbers at the top of the figure show the total number of genomes available for each influenza season in the primary dataset comprised of 452 genomes from which the ratio was calculated, while the numbers in brackets correspond to numbers of sequences in the larger secondary genomes dataset.
Numbers at the bottom are influenza seasons from the 1987/1988 (87/88) season to the 2011/2012 season.
Yamagata lineage of PA, NP, NA and MP segments and Victoria lineage of the NS segment eventually become fixed (in the population genetics sense of the word) in the influenza B population.
PB1, PB2 and HA segments maintain separate Victoria and Yamagata lineages.}
	\label{lineageRatiosOverTime}
\end{figure}

We show the ratio of Vic to Yam sequences in our primary and secondary datasets in different influenza seasons in Figure \ref{lineageRatiosOverTime}, which is based on which lineage each sequence was assigned to (see Methods).
It is evident that losses of diversity in the PA, NP, NA, MP and NS segments are related to either the Vic (NS) or the Yam (PA, NP, NA, MP) lineage replacing the other lineage in the influenza B virus population.
Similarly, the lack of reassortment between Vic and Yam lineages and maintenance of diversity of PB1, PB2 and HA can be seen, where the two lineages have been sequenced at a ratio close to 50\% over long periods of time (Figure \ref{lineageRatiosOverTime}).
On a year-to-year basis, however, the ratios for Vic and Yam sequences PB1, PB2 and HA can fluctuate dramatically consistent with one lineage predominating within a given season, in agreement with surveillance data \citep{reed2012}.

\begin{figure}[h]
	\centering		
	\includegraphics[width=0.95\textwidth]{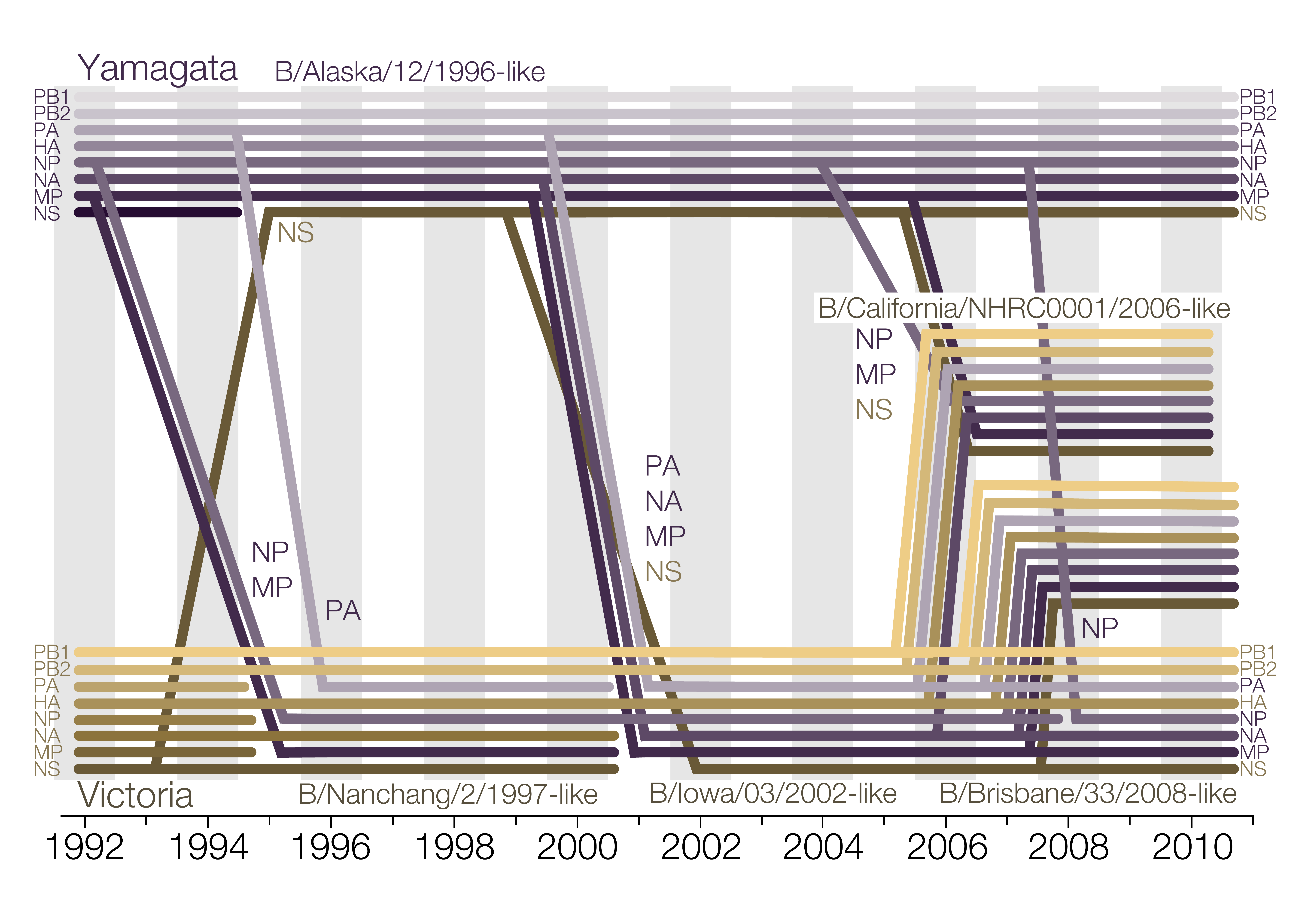}
	\caption{\textbf{Schematic plot of reconstructed reassortments between Victoria and Yamagata lineage segments of influenza B virus.}
Lineages that coassort in genomes are represented by 8 parallel lines, with lineages that derive from the original Victoria clade colored yellow/brown and lineages that derive from the original Yamagata clade colored lilac/purple.
Inter-lineage reassortment events are indicated by lines entering a different genome.
The angle of incoming lineages represents uncertainty in the timing of the event (mean date of the reassortant node and its parent node).
Lineage extinction dates are not shown accurately.}
	\label{railroadPlot}
\end{figure}

We reconstructed reassortment events that were detected by using lineage labels.
Figure \ref{railroadPlot} focuses only on inter-lineage reassortments that have occurred after 1990.
We identify 5 major (in terms of persistence) reassortant genome constellations (given in order PB1-PB2-PA-HA-NP-NA-MP-NS with prime (') indicating independently acquired segments) circulating between 1992 and 2011 (Figure \ref{railroadPlot}):
\begin{itemize}
  \item B/Alaska/12/1996-like (Y-Y-Y-Y-Y-Y-Y-V)
  \item B/Nanchang/2/1997-like (V-V-Y-V-Y-V-Y-V)
  \item B/Iowa/03/2002-like (V-V-Y'-V-Y-Y-Y'-V')
  \item B/California/NHRC0001/2006-like (V-V-Y-V-Y'-Y-Y'-V')
  \item B/Brisbane/33/2008-like (V-V-Y-V-Y'-Y-Y-V)
\end{itemize}
In a previous study B/Alaska/12/1996-like, B/Nanchang/2/1997-like and B/Iowa/03/2002-like constellations were observed \citet{chen2008}, but sequences from \\B/California/NHRC0001/2006-like and B/Brisbane/33/2008-like constellations were not available at the time.
In their study \citet{chen2008} also recovered the co-assortment pattern of PB1, PB2 and HA lineages, but did not remark upon it.
Of these 5 constellations 4 (B/Nanchang/2/1997-like, B/Iowa/03/2002-like,\\ B/California/NHRC0001/2006-like and B/Brisbane/33/2008-like) are derived from introgression of Yamagata lineage segments into Victoria lineage PB1-PB2-HA background, with only 1 (B/Alaska/12/1996-like) resulting from introgression of Victoria lineage NS segment into an entirely Yamagata derived background.
All 5 inter-lineage reassortment events described here are marked by the preservation of either entirely Victoria or Yamagata derived PB1-PB2-HA segments.
Figure \ref{railroadPlot} also shows that reassorting segments appear to evolve with a considerable degree of autonomy.
For example, the NP lineage that entered a largely Victoria lineage derived genome and gave rise to the B/Nanchang/2/1997-like isolates continued circulating until 2010, even though the other segments it co-assorted with in 1995 -- 1996 (PA and MP) went extinct following the next round of reassortment that led to the rise of B/Iowa/03/2002-like genome constellations.
A more extreme example is the NS segment, where a Vic sub-lineage was reassorted into an entirely Yam background (B/Alaska/12/1996-like) in 1994--1995, then reassorted back into a mostly Vic background some 5 years later (B/Iowa/03/2002-like) where it has replaced the `original' Vic sub-lineage (see Figure \ref{railroadPlot}).

We observe that in all 5 successful inter-lineage reassortment events shown in Figure \ref{railroadPlot}, none break up the PB1-PB2-HA complex.
This is an unlikely outcome -- the probability of not breaking up PB1-PB2-HA across 5 reassortment events is $p = (\frac{2^{5} \times 2 - 2}{2^{8} - 2})^{5} = 0.0009$, where reassortment events are considered to sample from the Vic and Yam lineages at random for each of the 8 segments.
If we correct for multiple testing with the assumption that co-assortment of any 3 segments is of interest we find that the probability of not breaking up an arbitrary set of 3 segments across 5 reassortment events is $p = \binom{8}{3} \times (\frac{2^{5} \times 2 - 2}{2^{8} - 2})^{5} = 0.0485$.

\begin{figure}[h]
	\centering		
	\includegraphics[width=0.65\textwidth]{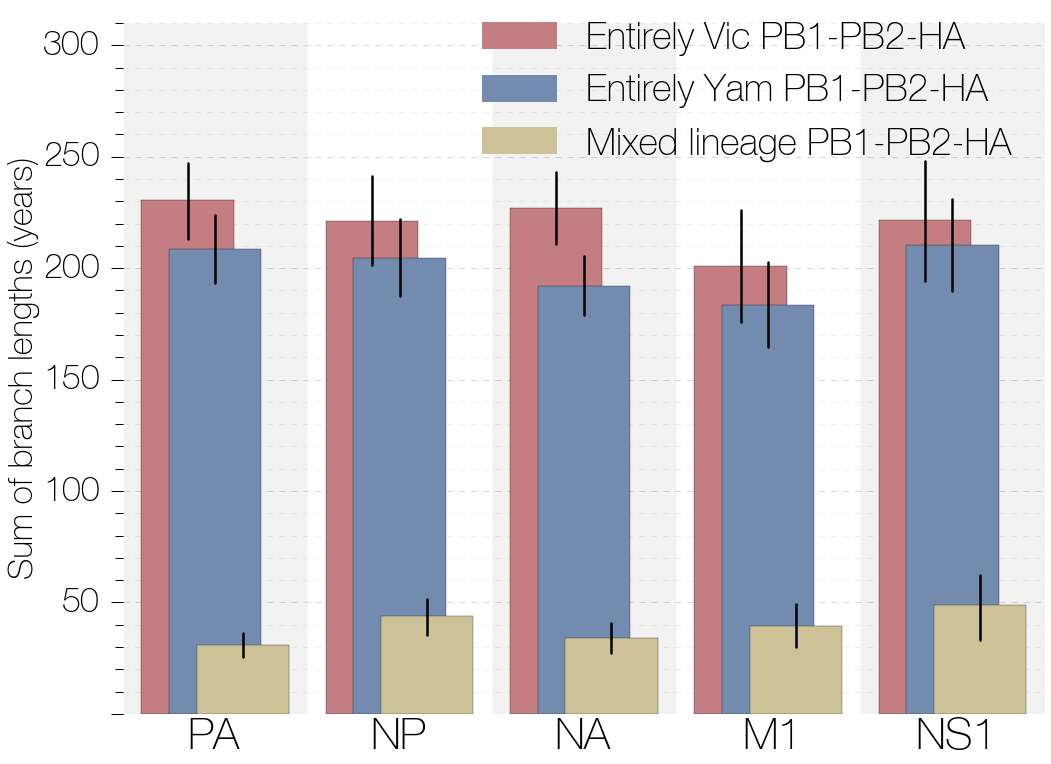}
	\caption{\textbf{Amount of evolutionary time each segment has spent under different PB1-PB2-HA constellations.}
All segments have spent significantly more of their history with entirely Vic or entirely Yam-derived PB1-PB2-HA complexes.
All vertical lines indicating uncertainty are 95\% highest posterior densities (HPDs).}
	\label{stateTime}
\end{figure}

Although the vast majority of influenza B isolates possess either Vic or Yam lineage derived PB1-PB2-HA complexes, on rare occasions mixed-lineage PB1-PB2-HA constellations emerge.
Figure \ref{stateTime} shows the sum of branch lengths which were labelled as having entirely Vic, entirely Yam or mixed-lineage PB1, PB2 and HA segments.
Due to lack of reassortment between Vic and Yam lineages of PB1, PB2 and HA (Figure \ref{betweenDiversity}) since 1997 all segments have spent significantly longer periods of evolutionary time with either entirely Vic-derived or entirely Yam-derived than with mixed-lineage PB1, PB2 and HA constellations (Figure \ref{stateTime}).
We have identified 3 instances of mixed-lineage PB1-PB2-HA reassortants from the primary dataset with the following PB1-PB2-HA constellations: VVY (B/Bangkok/163/1990-like, 13 sequences isolated 1990 -- 5 Jan 1995), VYV (B/Nanchang/630/1994-like, 2 sequences isolated 1994 -- 1996) and VYY (B/New York/24/1993-like, 2 sequences isolated 8 Jan 1993 -- 1994).
We detected two new reassortant lineages when investigating the larger secondary dataset -- B/Waikato/6/2005-like viruses with PB1-PB2-HA constellation YYV (17 sequences isolated 9 May -- 12 October in 2005) and B/Malaysia/1829782/2007 with PB1-PB2-HA constellation YVY (1 sequence isolated 2 August 2007).

\subsection*{Analysis of reassortment properties}
\begin{figure}[h]
	\centering
	\includegraphics[width=0.65\textwidth]{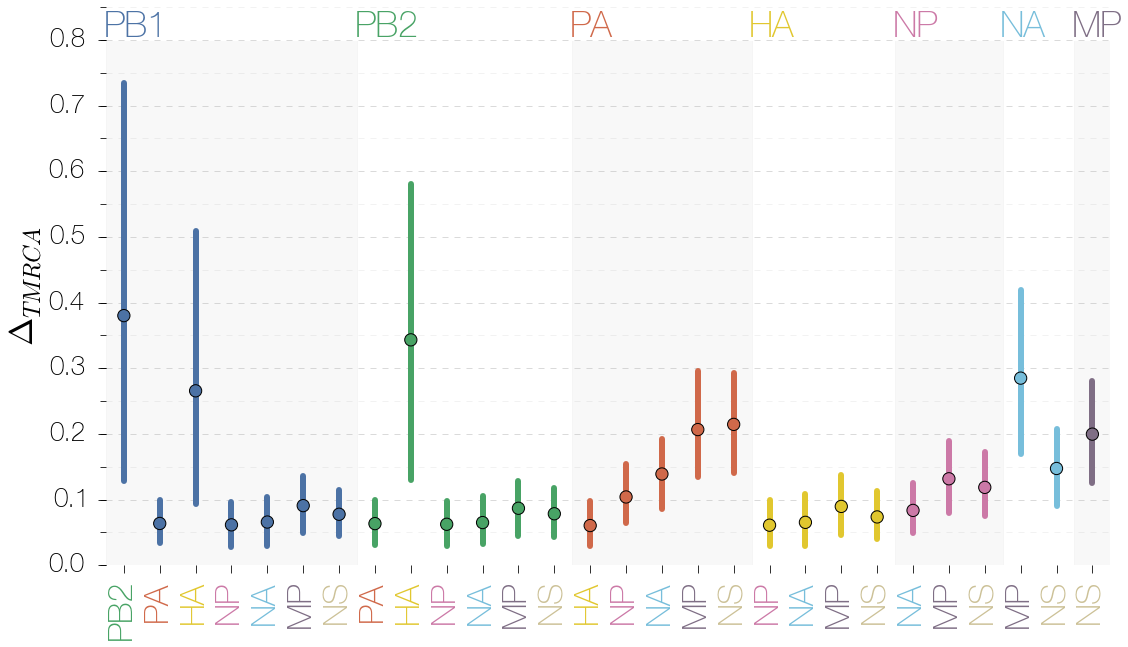}
	\caption{\textbf{$\dtmrca$ statistics for different segment pairs.}
PB1, PB2 and HA trees exhibit reciprocally highly similar TMRCAs, unlike most other pairwise comparisons.
All vertical lines indicating uncertainty are 95\% highest posterior densities (HPDs).}
	\label{deltaTMRCA}
\end{figure}

We attempted to quantify the temporal discordance between lineages reassorting into new genomic constellations.
If one were able to recover an influenza `species tree', including admixture/reassortment events, it would be possible to estimate the reassortment or recombination `distance', which is the time between a split in the species tree in the past and a reassortment event (see Figure \ref{speciesTree}).
Although we do not find evidence of differences in total number of reassortments between segments (see Supplementary information), we find support for a reassortment `distance' effect, in which a pair of tips on one segment has a different TMRCA from the same pair of tips on a different segment.
The summary statistic we use that reflects this difference in TMRCAs, $\undtmrca$, is most sensitive when only one of the two trees being compared loses diversity via reassortment and the other acts like a proxy for the `species tree'.
We normalize our $\undtmrca$ comparisons to arrive at $\dtmrca$, which accounts for uncertainty in tree topology (see Methods).
Figure \ref{deltaTMRCA} shows $\dtmrca$ values for all pairs of trees.
Most segment pairs show very low values for this statistic with $\dtmrca \approx 0.1$, indicating that $\undtmrca$ measurements between replicate posterior samples from the same segment are up to 10 times smaller than $\undtmrca$ values between posterior samples from different segments.
PB1, PB2 and HA trees, on the other hand, exhibit $\dtmrca$ values that are much higher.
This shows that TMRCA differences between trees of PB1, PB2 and HA segments are, though noisy, occasionally very similar to uncertainty in tip-to-tip TMRCAs between replicate analyses of these segments.

\section*{Discussion}

\subsection*{Linkage between PB1, PB2 and HA gene segments}
In this paper we show that the PB1, PB2 and HA segments of influenza B viruses are the only ones that have continuously maintained separate Vic and Yam lineages, while other segments have fixed either Vic or Yam lineages (Figures \ref{genomeGrid}, \ref{lineageRatiosOverTime} and \ref{railroadPlot}).
Evidence suggests that this is a result of prolonged lack of reassortment between Vic and Yam lineages in PB1, PB2 and HA (Figure \ref{betweenDiversity}) which possess co-assorting sequences detectable as high linkage disequilibrium (Figure \ref{segmentLD}).
The vast majority of the sampled evolutionary history of each segment of influenza B viruses since the split of Vic and Yam lineages has been spent in association with either completely Victoria or completely Yamagata lineage derived PB1-PB2-HA complexes (Figure \ref{stateTime}), suggesting that having `pure' lineage PB1-PB2-HA complexes is important for whole-genome fitness.
We propose that this pattern of coassortment is due to the action of selection and not simply biased or rare reassortment.

The origin of the strong genetic linkage between PB1, PB2 and HA segments remains unclear.
We believe there are two alternative, but similar explanations for the origins of the strong genetic linkage between these segments: mutation-driven co-evolution \citep{presgraves2010} and Dobzhansky-Muller incompatibility \citep{dobzhansky1937,muller1942}.
Mutation-driven co-evolution \citep{presgraves2010}, has been suggested to be the cause of hybrid dysfunction in \textit{Saccharomyces} hybrids \citep{lee2008}, and evolves as a by-product of adaptation.
If one or the other influenza B lineage has undergone adaptation we might expect these changes to be beneficial in its native background and incompatible with a foreign background.
Dobzhansky-Muller incompatibility operates in a similar way, but the main difference from the scenario described earlier is that the incompatible alleles are neutral or nearly neutral in their native background and become deleterious or lethal when combined with non-native backgrounds.
Emergence of DM incompatibility is aided by geographic isolation.
Interestingly, the Victoria lineage of HA was restricted to eastern Asia between 1992 and 2000 \citep{nerome1998,shaw2002}, offering ample time for the budding Victoria lineage to accumulate alleles causing reassortment incompatibility.
However, without more genomic data from the past, it is difficult to estimate to what extent influenza B virus population structure contributed to the development of the current segment linkage.

\subsection*{Potential mechanisms for reassortment incompatibility}
Unfortunately, the limited amount of genomic data available for the early years of the Vic--Yam split precludes any attempts of answering whether selection or drift have led to the current linkage of PB1, PB2 and HA segments.
Although the origins of the linkage between these three segments might be difficult to explain, we can speculate on the nature of reassortment incompatibility.
For example, it is intuitive for why this might be the case for PB1 and PB2: both proteins interact directly as part of the RNA-dependent RNA polymerase heterotrimer.
Indeed, we observe that PB1--PB2 reassortants are the rarest and least persistent among mixed-lineage PB1-PB2-HA strains and have not been isolated in great numbers.
In fact most reassortants breaking the PB1-PB2-HA complex apart have occured in the past, close to the split of Vic and Yam lineages and have become very rare since.

There is some evidence that the linkage between PB1 and HA might not be a phenomenon restricted to influenza B viruses.
It has been established that at least for the 1957 and the 1968 influenza pandemics, caused by A/H2N2 and A/H3N2 subtypes, respectively, the viruses responsible were reassortants possessing PB1 and HA segments derived from avian influenza A viruses \citep{kawaoka1989}.
In addition, outdated techniques for producing vaccine seed strains through selection for HA-NA reassortants often yielded PB1-HA-NA reassortants as a side-effect \citep{bergeron2010,fulvini2011}.
Recent experiments have found that the presence or absence of a `foreign' PB1 segment can have dramatic effects on HA concentration on the surface of virions and total virion production \citep{cobbin2013}.
However, there have been reassortant influenza A viruses circulating for prolonged periods of time in humans that did have disparate PB1 and HA segments, \textit{e.g.} H1N2 outbreaks in 2001 \citep{gregory2002} and H1N1/09 in 2009 \citep{smith2009}.

We believe that the association between PB1, PB2 and HA segments should be relatively straightforward to explore in the lab.
Reverse genetics systems have been developed for influenza B viruses \citep{hoffmann2002}, which would allow the creation of artificial reassortants.
Based on the frequency and persistence times of different reassortant classes we have observed, we expect a hierarchy of reassortant fitness starting with PB1+PB2+HA reassortants which should be the most fit, followed by PB1+2/HA, then PB1+HA/PB2 and finally PB2+HA/PB1 reassortants with the lowest fitness.
We believe that this is the most direct approach to unravelling the mechanism responsible for the linkage within the PB1-PB2-HA complex.

\subsection*{Will influenza B viruses speciate?}
We suggest that the preservation of two PB1-PB2-HA complex lineages is similar to genomic speciation islands, where small numbers of genes resist being homogenized through gene flow \citep{turner2005}.
In this context, we see three potential paths of evolution for influenza B viruses.
If more segments get recruited to the PB1-PB2-HA complex, the process could continue until `speciation' occurs in which none of the segments are able to reassort across the Victoria--Yamagata lineage boundary.
Alternatively, the influenza B genome could continue to be homogenized via gene flow with the exception of PB1, PB2 and HA segments or one of the two PB1-PB2-HA complexes could go extinct, marking the return of single-strain dynamics in the influenza B virus population.
The eventual fate of influenza B viruses will likely be determined by the combined effects of reassortment frequency and the strength of epistatic interactions between segments.

\newpage

\section*{Methods}
We compiled a primary dataset of 452 complete influenza B genomes from GISAID \citep{GISAID} dating from 1984 to 2012.
The longest protein coding region of each segment was extracted and used for all further analyses.
We thus assume that homologous recombination has not taken place and that the evolutionary history of the whole segment can be inferred from the longest coding sequence in the segment.
To date there has been little evidence of homologous recombination in influenza viruses \citep{chare2003,boni2008,han2010}.
The segments of each strain were assigned to either Vic or Yam lineage by making maximum likelihood trees of each segment using PhyML \citep{guindon2003} and identifying whether the isolate was more closely related to B/Victoria/2/87 or B/Yamagata/16/88 sequences in that segment, with the exception of the NS segment since B/Victoria/2/87 was a reassortant and possessed a Yam lineage NS \citep{lindstrom1999}.
B/Czechoslovakia/69/1990 was considered as being representative of Victoria lineage for the NS segment.
Every segment in each genome thus received either a Vic or a Yam lineage designation, for example the strain B/Victoria/2/87 received V-V-V-V-V-V-V-Y, since its NS segment is derived from the Yam lineage and the rest of the genome is Vic.

We also collated a secondary dataset from all complete influenza B virus genomes available on GenBank as of May 5, 2014.
After removing isolates that had considerable portions of any sequence missing, were isolated prior to 1980 or were suspected of having a contaminant sequence in any segment, we were left with 1603 sequences.
This dataset only became available after all primary analyses were performed, are mainly from Australia, New Zealand and the United States and are too numerous to analyze in BEAST \citep{drummond2012}.
PhyML \citep{guindon2003} was used to produce phylogenies of each segment and the lineage of each isolate was determined based on grouping with either B/Victoria/2/87 or B/Yamagata/16/88 sequences, as described above.
By associating strains with lineage identity of each of their segments, we reconstructed the most parsimonious inter-lineage reassortment history for the secondary dataset.
The secondary dataset was used to check how representative the primary dataset was, to estimate LD and to broadly confirm our results.
All analyses pertain to the primary dataset unless stated otherwise.

Temporally-calibrated phylogenies were recovered for each segment in the primary dataset using Markov chain Monte Carlo (MCMC) methods in the BEAST software package \citep{drummond2012}.
We modeled the substitution process using the HKY model of nucleotide substitution \citep{hky1985}, with separate transition models for each of the 3 codon partitions, and additionally estimated realized synonymous and non-synonymous substitution counts \citep{obrien2009}.
We used a flexible Bayesian skyride demographic model \citep{minin2008}.
We accounted for incomplete sampling dates for 94 sequences (of which 93 had only year and 1 had only year and month of isolation) whereby tip date is estimated as a latent variable in the MCMC integration.
A relaxed molecular clock was used, where branch rates are drawn from a lognormal distribution \citep{drummond2006}.
We ran 3 independent MCMC chains, each with 200 million states, sampled every 20,000 steps and discarded the first 10\% of the MCMC states as burn-in.
After assessing convergence of all 3 MCMC chains by visual inspection using Tracer \citep{tracer}, we combined samples across chains to give a total of 27,000 samples from the posterior distribution of trees.

Every sequence was assigned 7 discrete traits in BEAUti corresponding to the lineages of all other segments with which a strain was isolated \textit{e.g.} PB1 tree had PB2, PA, HA, NP, NA, MP and NS as traits and V or Y as values for each trait.
We inferred the ancestral state of lineages in each segment by modelling transitions between these discrete states using an asymmetric transition matrix \citep{lemey2009} with Bayesian stochastic search variable selection (BSSVS) to estimate significant rates.
Because the posterior set of trees for a single segment has branches labelled with the inferred lineage in the remaining 7 segments, we can detect inter-lineage reassortments between pairs of segments by observing state transitions, i.e.\ Yam to Vic or Vic to Yam (Figure \ref{methodFig}).
In addition, by reconstructing the ancestral state of all other genomic segments jointly we can infer co-reassortment events when more than one trait transition occurs on the same node in a tree.
Inter-phylogeny labeling approaches have been extensively used in the past to investigate reticulate evolution in influenza A viruses and HIV \citep{lycett2012,ward2013,lu2014}.

\subsection*{Measures of diversity}
We inferred the diversity of each segment from their phylogenetic tree by estimating the date of the most recent common ancestor of all branches at yearly time points, which places an upper bound on the maximum amount of diversity existing at each time point.
A version of this lineage turnover metric has previously been used to investigate the tempo and strength of selection in influenza A viruses during seasonal circulation \citep{bedford2011}.
In addition, we calculated mean pairwise time of most recent common ancestor (TMRCA) between branches labelled as Vic and Yam for PB1, PB2 and HA traits.
This gave us a measure of how much a particular segment reassorts with respect to Vic and Yam lineages of PB1, PB2 and HA segments.
If Vic and Yam lineages of PB1, PB2 and HA segments were to be considered as being separate populations this measure would be equivalent to `between population' diversity.

We also calculated the total amount of sampled evolutionary time spent by each segment with entirely Vic, entirely Yam or mixed lineage PB1, PB2 and HA segments.
We do this by summing the branch lengths in each tree under 3 different lineage combinations of the PB1, PB2 and HA segments: PB1-PB2-HA derived entirely from Yamagata lineage, PB1-PB2-HA entirely derived from Victoria lineage and PB1-PB2-HA derived from a mixture of the two lineages.
This gives a measure of how successful, over long periods of time, each particular PB1-PB2-HA constellation has been.

\subsection*{Tree to tree similarities}
We express the normalized distance $\dtmrca$ between trees belonging to two segments $A$ and $B$ for a particular posterior sample $i$, following
\begin{equation}
\dtmrca(A_i, B_i) = \frac{\undtmrca(A_i, A_i') + \undtmrca(B_i, B_i')}{2 \, \undtmrca(A_i, B_i)},
\end{equation}
where $\undtmrca(A_i, B_i) = \frac{1}{n}\sum_{j=1}^n g(A_{ij}, B_{ij})$ and $n$ is the total number of pairwise comparisons available between sets of tips.
Thus, $g(A_{ij},B_{ij})$ is the absolute difference in TMRCA of a pair of tips $j$, where the pair is drawn from the \textit{i}th posterior sample of tree $A$ and the \textit{i}th posterior sample of tree $B$.
Additionally, $\undtmrca(A_i,A'_i)$ is calculated from the \textit{i}th posterior sample of tree A and \textit{i}th posterior sample of an independent analysis of tree $A$ (which we refer to as $A'$), which is used in the normalization procedure to control for variability in tree topology stability over the course of the MCMC chain (see Figures \ref{deltaTMRCAtrees} and \ref{deltaTMRCAreplicates}).
We had 3 replicate analyses of each segment and in order to calculate $\undtmrca(A_i,A'_i)$ we used analyses numbered 1, 2 and 3 as $A$ and analyses numbered 2, 3 and 1 as $A'$, in that order.
We subsampled our combined posterior distribution of trees to give a total of 2,700 trees on which to analyze $\dtmrca$.

Calculating the normalized $\dtmrca(A_i, B_i)$ for each MCMC state provides us with a posterior distribution of this statistic allowing  specific hypotheses regarding similarities between the trees of different segments to be tested.
Our approach exploits the branch scaling used by BEAST \citep{drummond2012}, since the trees are scaled in absolute time and insensitive to variation in nucleotide substitution rates between segments, allowing for direct comparisons between TMRCAs in different trees.
In the absence of reassortment we expect the tree of every segment to recapitulate the `virus tree', a concept analogous to `species trees' in population genetics.
Our method operates under the assumption that the segment trees capture this `virus tree' of influenza B viruses quite well.
It is not an unreasonable assumption, given the seasonal bottlenecks influenza viruses experience.
This makes it almost certain that influenza viruses circulating at any given time point are derived from a single genome that existed in the recent past.
The $\undtmrca$ statistic essentially quantifies the temporal distance between admixture events and nodes in the `virus tree' (see Figure \ref{speciesTree}).
We normalize $\undtmrca$ values to get $\dtmrca$, a measure which quantifies the extent to which the similarity of two independent trees resembles phylogenetic noise.
The $\undtmrca$ statistic is an extension of patristic distance methods and has previously been used to tackle a wide variety of problems, as phylogenetic distance in predicting viral titer in \textit{Drosophila} infected with viruses from closely related species \citep{longdon2011} and to assess temporal incongruence in a phylogenetic tree of amphibian species induced by using different calibrations \citep{ruane2011}.

\subsection*{Linkage disequilibrium across the influenza B genome}
We used the secondary GenBank dataset with 1603 complete genome sequences to estimate linkage disequilibrium (LD) between amino acid loci across the longest proteins encoded by each segment of the influenza B virus genome.
To quantify LD we adapt the $\chiSq$ statistic from \citep{hedrick1986}:

\begin{equation}
\chiSq=\frac{\chi^{2}}{N\,(k-1)\,(m-1)},
\end{equation}

where $\chi^{2}$ is calculated from a classical contingency table, $N$ is the number of haplotypes and $(k-1)(m-1)$ are the degrees of freedom.
This statistic is equal to the widely used $r^2$ LD statistic at biallelic loci, but also quantifies LD when there are more than two alleles per locus \citep{zhao2005}.
LD was estimated only at loci where each nucleotide or amino acid allele was present in at least two isolates.
We ignored gaps in the alignment and did not consider them as polymorphisms.
In all cases we used a minor allele frequency cutoff of 1\%.
We also calculated another LD statistic, $D'$ \citep{lewontin1964} as $D'_{ij}=D_{ij}/D_{ij}^{max}$,
where $D_{ij}=p(A_i B_j) - p(A_i) p(B_j)$ and
\begin{equation}
\begin{split}
D_{ij}^{max}=min[ p(A_i) p(B_j) , (1-p(A_i))(1-p(B_j)) ] \mbox{ when } D_{ij} < 0\\
D_{ij}^{max}=min[ (1-p(A_i)) p(B_j) , p(A_i)(1-p(B_j)) ] \mbox{ when } D_{ij} \geq 0,
\end{split}
\end{equation}
where $p(A_i)$ is the frequency of allele $A_i$ at locus A, $p(B_j)$ is the frequency of allele $B_j$ at locus B and $p(A_i B_j)$ is the frequency of haplotype $A_i B_j$.
$D'$ is inflated when some haplotypes are not observed \textit{e.g.} when the minor allele frequency is low.
We find that $D'$ is almost uniformly high across the influenza B virus genome and close to 1.0 for almost any pair of polymorphic loci.
This is because most amino acid alleles in the population exist transiently, meaning that they do not get a chance to reassort and we only observe them within the backgrounds of more persistent alleles, which $D'$ quantifies as complete LD.
We think that metrics related to $r^2$, like $\chiSq$, perform much better on temporal data such as ours in finding persistent associations between alleles and are easier to interpret.

\subsection*{Data availability}
Python scripts used to process trees and sequences are available at:
\\https://github.com/evogytis/fluB/tree/master/scripts.

Output files from scripts, lineage designations, MCC trees, acknowledgment tables, accession numbers and redacted XML files (per GISAID Data Access Agreement) are publicly available at:
\\https://github.com/evogytis/fluB/tree/master/data.

\section*{Acknowledgements}
We would like to thank Darren Obbard and Paul Wikramaratna for helpful discussions and anonymous reviewers for comments and suggestions.
GD was supported by a Natural Environment Research Council studentship D76739X.
TB was supported by a Newton International Fellowship from the Royal Society. 
The research leading to these results has received funding from the European Research Council under the European Community's Seventh Framework Programme (FP7/2007-2013) under Grant Agreement no. 278433-PREDEMICS and ERC Grant agreement no. 260864.
AR and SL acknowledge the support of the Wellcome Trust (grant no. 092807).




\clearpage

\title{\vspace{1.0cm} \huge \bf Supplemental information:\\ \LARGE Reassortment between influenza B lineages and the emergence of a co-adapted PB1-PB2-HA gene complex}



\date{\today}

\maketitle

\setcounter{figure}{0}
\setcounter{table}{0}
\renewcommand{\thefigure}{S\arabic{figure}}
\renewcommand{\thetable}{S\arabic{table}}

\subsection*{Confirmation of primary findings}
\begin{figure}[h]
	\centering	
	\includegraphics[width=0.85\textwidth]{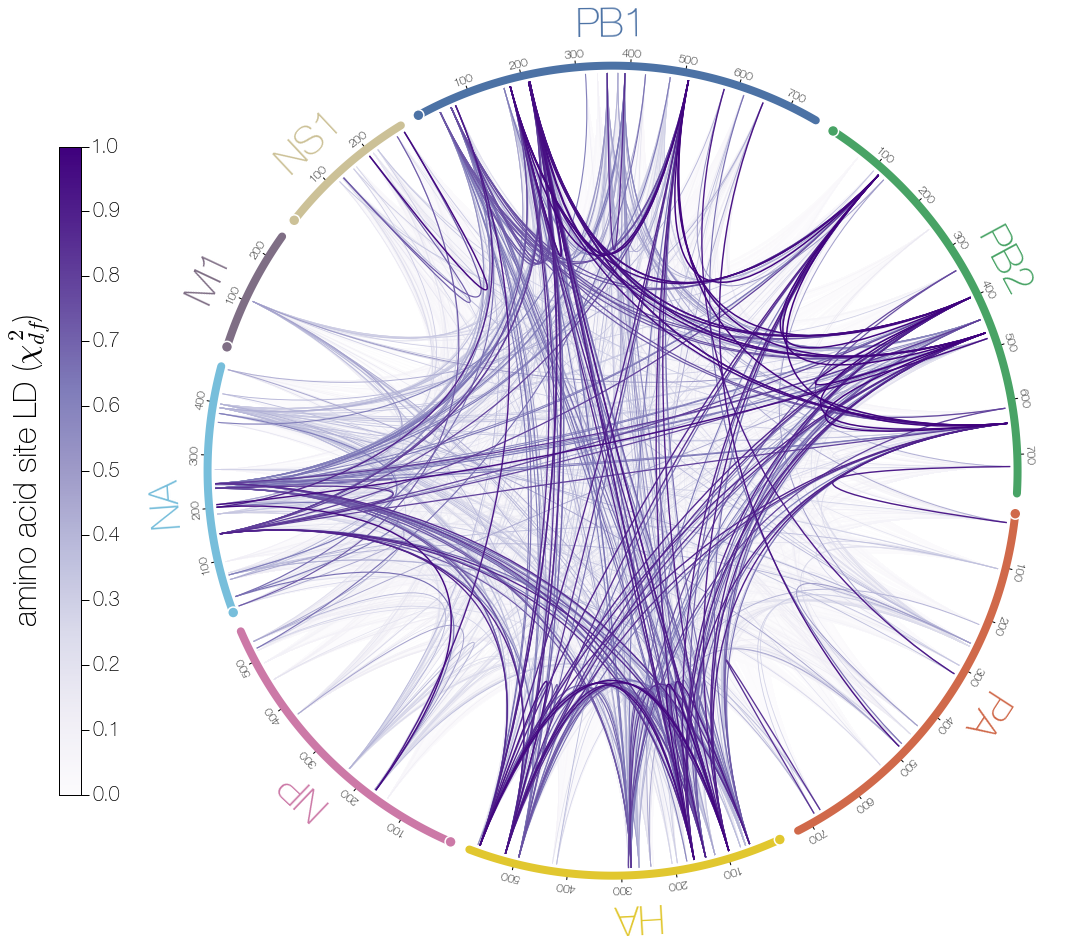}
	\caption{\textbf{LD comparison between influenza B proteins.}
Pairwise comparisons of linkage disequilibrium between amino acid sites on influenza B proteins in the secondary dataset.
Many polymorphic amino acid sites on PB1, PB2 and HA proteins exhibit high LD between themselves, followed by the NA protein.
This is evidence of a considerable number of co-assorting alleles within these proteins.}
	\label{segmentLD}
\end{figure}

\begin{figure}
\centering  
\includegraphics[width=0.85\textwidth]  {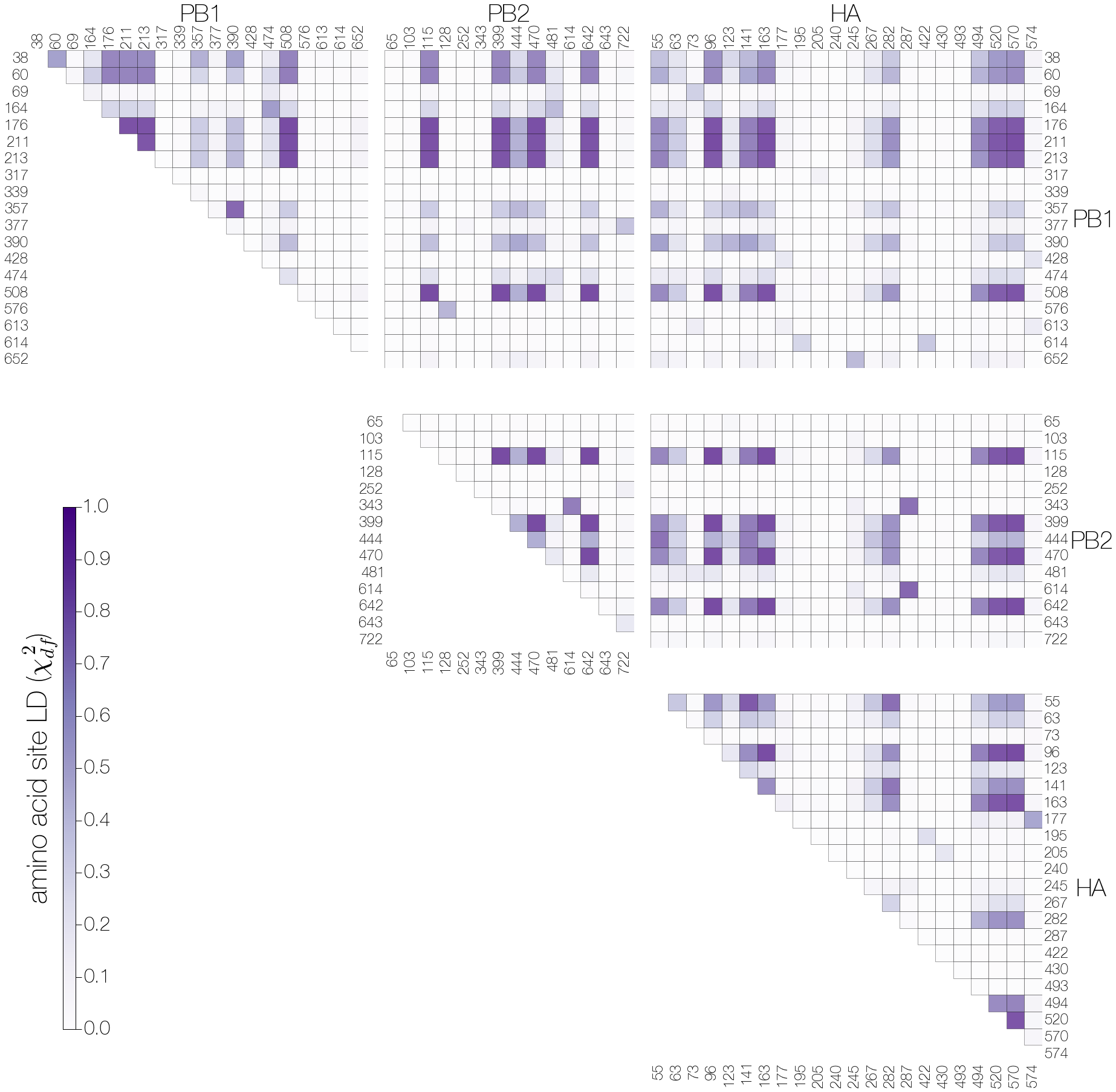}
\caption{\textbf{Heatmap of linkage disequilibrium ($\chiSq$) between amino acid sites on PB1, PB2 and HA proteins.}
Numbers next to each row and column correspond to amino acid site number within a given protein starting from methionine.
Amino acid sites exhibiting reciprocally high LD between PB1, PB2 and HA proteins are: 176, 211, 213, 508 (PB1), 115, 399, 470, 642 (PB2) and 96, 163, 520 and 570 (HA).
Sites 211 and 213 on the PB1 protein are very close to each other and the stretch of sequence around these residues contains many positively charged amino acids (lysine and arginine).
Multiple nuclear localization signals (NLSs) are predicted to occur around this region and sites 211 and 213 are either predicted to be near the end of a mono-partite NLS or the beginning of a bi-partite NLS.
Previous research \citep{nath1990} suggests that in the influenza A PB1 protein residue 211 (homologous to influenza B PB1 residue 211) is the last residue of a bi-partite NLS.
Almost all Yamagata lineage isolates possess arginine (R) residue at PB1 position 211 and a serine (S) residue at position 213, whereas Victoria lineage isolates have lysine (K) at position 211 and threonine (T) at position 213.
It remains to be seen whether these sites significantly affect the nuclear import efficiency of the PB1 protein of either lineage.
Though the PB1 protein is known to accumulate in the nucleus on its own, efficient import into the nucleus requires the presence of the PA protein \citep{fodor2004}.
Similarly, site 399 on the PB2 protein are close to residues 377, 406 and 408 which are homologous to sites in influenza A that are responsible for mRNA cap-binding \citep{guilligay2008}.}
\label{ChiCore}
\end{figure}

\begin{figure}
\centering  
\includegraphics[width=0.95\textwidth]  {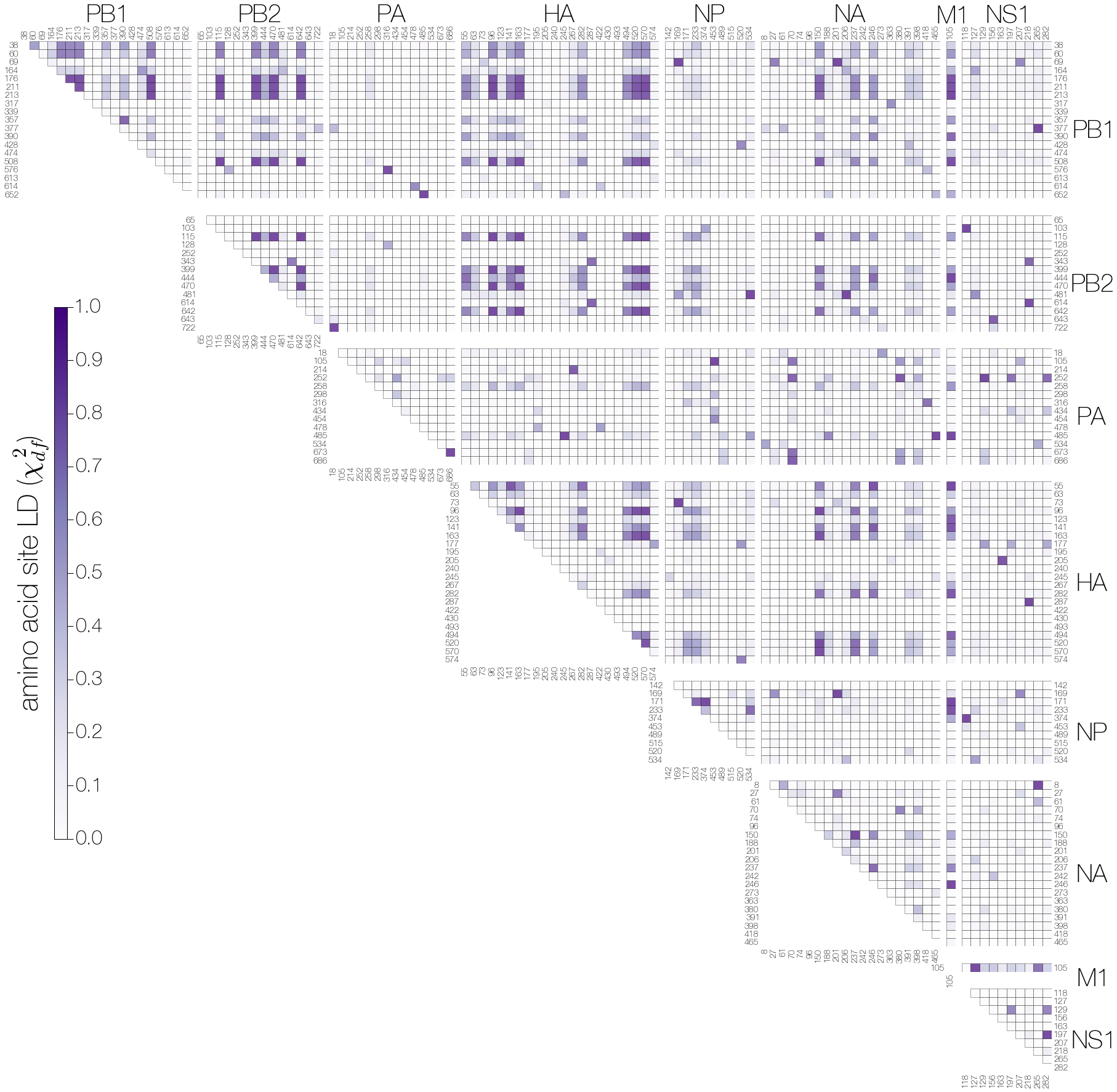}
\caption{\textbf{Heatmap of genome-wide linkage disequilibrium ($\chiSq$) between polymorphic amino acid sites.}
Patterns of LD across the genome suggest a network of reciprocally linked amino acid sites on PB1, PB2, HA and, to some extent NA, proteins.
Proximity of sites on heatmaps might not correspond to proximity of sites within proteins.}
\label{ChiGenome}
\end{figure}

We sought to confirm our findings through measurement of linkage disequilibrium (LD), a measure of non-random association between polymorphic loci within a population.
We estimated LD directly from haplotype frequencies at polymorphic amino acid sites (see Methods) in the secondary dataset, thereby avoiding phylogenetic reconstruction or Vic/Yam lineage assignment.
We observe greater amino acid LD values between PB1, PB2 and HA than between other pairs of segments (Figure \ref{segmentLD}) in a large secondary dataset.
This suggests that PB1, PB2 and HA segments possess a considerable number of co-assorting non-synonymous alleles, which upon closer inspection are associated with either Vic or Yam lineage segments.
We conclude that Victoria and Yamagata lineages of PB1, PB2 and HA have accumulated lineage-specific amino acid substitutions.
Of the amino acid sites that exhibit high LD on PB1, PB2 and HA proteins, there are 4 sites on PB1, 4 on PB2 and 4 on HA proteins which form a network of sites exhibiting high LD (Figures \ref{ChiGenome} and \ref{ChiCore}).
These sites define the split between Vic and Yam lineages within PB1, PB2 and HA segments.
In addition, there are sites on PB1, PB2, HA and NA proteins which also show high, albeit smaller, LD which correspond to sites which have undergone amino acid replacements some time after the Vic/Yam split.

\subsection*{Analysis of within-lineage reassortment patterns}
\begin{figure}[h]
	\centering		
	\includegraphics[width=0.65\textwidth]{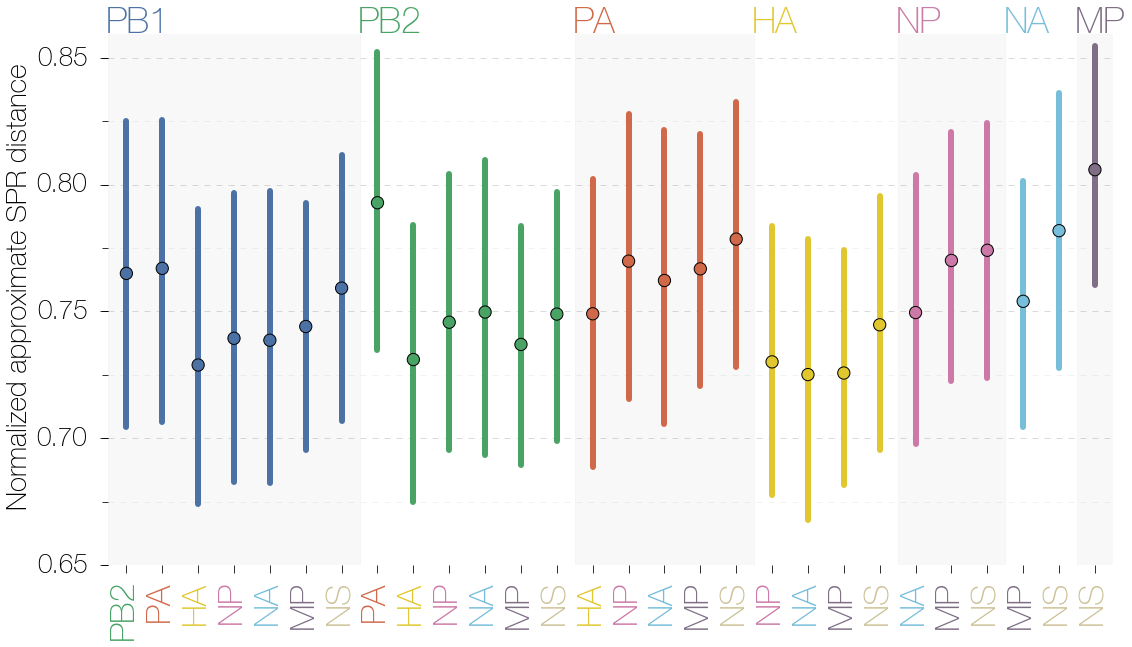}
	\caption{\textbf{Normalized approximate SPR distances between pairs of segments.}
Following the normalization procedure approximate SPR distances are similar across all pairwise comparisons.
We interpret this as lack of evidence for small-scale topological similarities between trees of all segments, which we expect to arise if any two segments were being co-packaged and co-reassorted.
All vertical lines indicating uncertainty are 95\% highest posterior densities (HPDs).}
	\label{SPRdistances}
\end{figure}

Subtree prune and regraft (SPR) distances between phylogenetic trees are an approximate measure of the numbers of reassortment or recombination events \citep{svinti2013}.
Exact SPR distances are difficult to compute, as they depend on the SPR distance itself and are impractical to compute for posterior distributions of trees except for the most similar trees.
We calculated approximate SPR distances \citep{whidden2009,whidden2010,whidden2013} to quantify the numbers of reassortments that have taken place between all pairs of segments.
Normalized approximate SPR distances,$\dspr$, were recovered using (see Methods):
\begin{equation}
\dspr(A_i, B_i) = \frac{f(A_i, A_i') + f(B_i, B_i')}{2 \, f(A_i, B_i)},
\end{equation}
where $f(A_i, A_i')$, $f(B_i, B_i')$ and $f(A_i, B_i)$ are approximate SPR distances between \textit{i}th posterior samples from segments $A$, $B$ and independent analyses thereof ($A'$ and $B'$).
Figure \ref{SPRdistances} shows approximate SPR distances between all pairs of segment trees after normalization.
If there are biases in the way segments reassort, so that some segments tend to co-assort more often, we expect to observe a lower reassortment rate between them, which would manifest as small-scale similarities between phylogenetic trees of those segments.
In our case we expect SPR distances, which are proportional to the number of reassortment events that have taken place between trees, to reflect the overall (\textit{i.e} both within and between lineages) reassortment rate.

\begin{figure}
\centering  
\includegraphics[width=0.65\textwidth]  {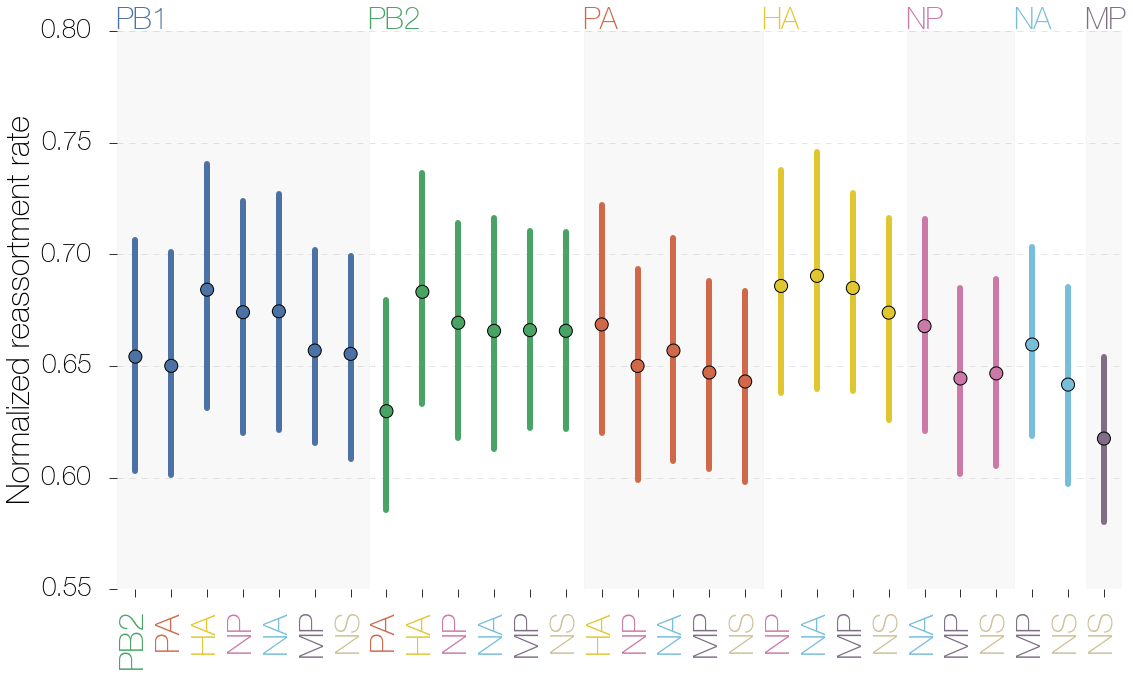}
\caption{\textbf{Normalized reassortment rate}
Reassortment rate is calculated as approximate number of SPR moves per sum of total time in both trees.}
\label{NormSPR_RErate}
\end{figure}

\begin{figure}
\centering  
\includegraphics[width=0.65\textwidth]  {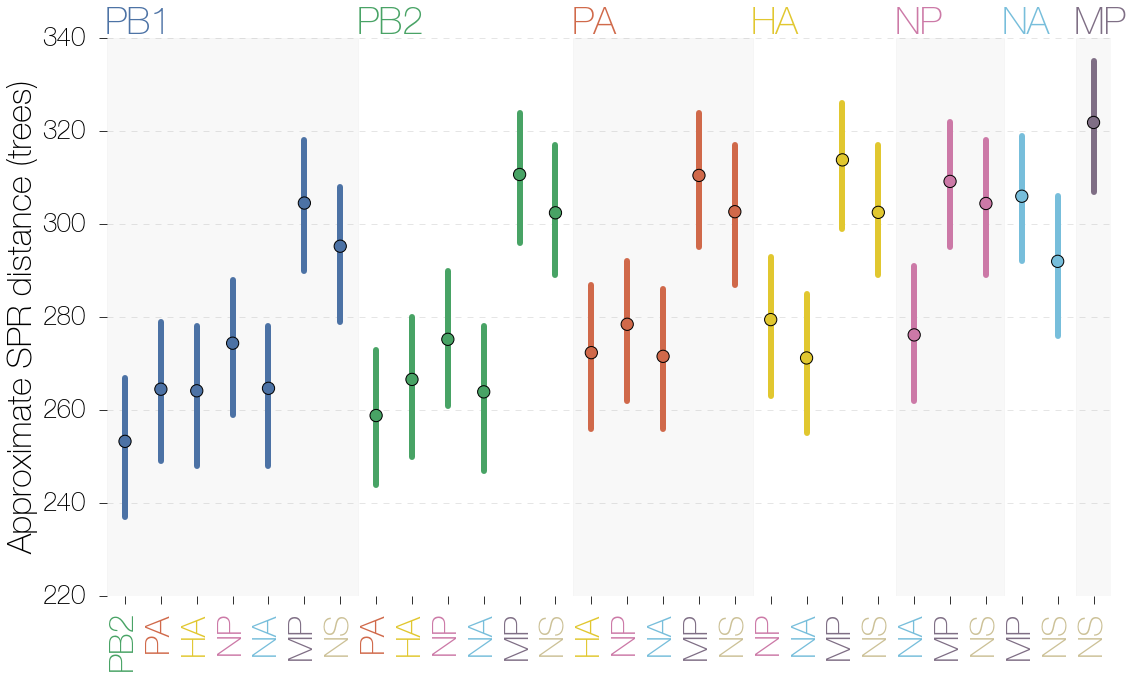}
\caption{\textbf{Approximate SPR distances between all pairs of trees of segments.}
There is a visible trend where comparisons between shorter segments have larger SPR distances, consistent with decreasing tree topology stability over the course of MCMC for shorter segments.}
\label{SPRdistancesTrees}
\end{figure}

\begin{figure}
\centering  
\includegraphics[width=0.65\textwidth]  {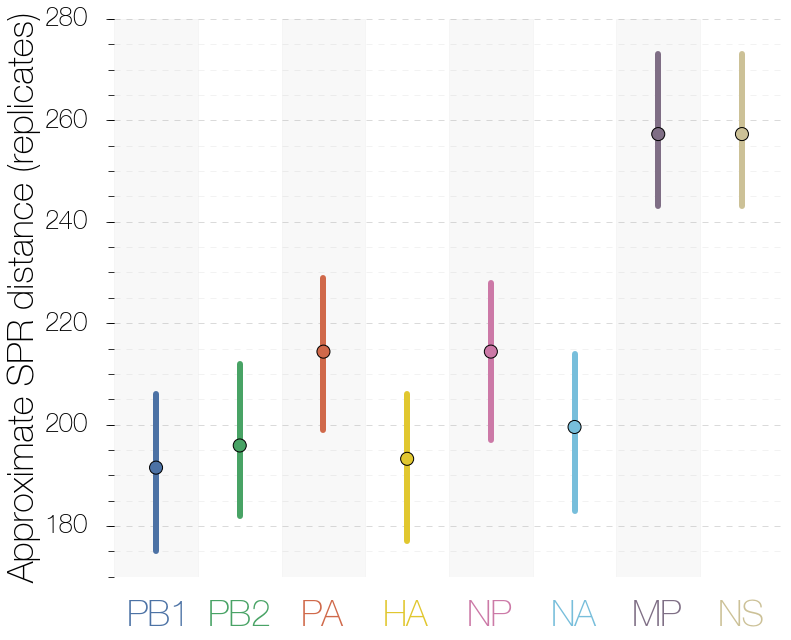}
\caption{\textbf{Approximate SPR distances between replicate trees of each segment.}
Approximate SPR distances between replicates of MP and NS trees are much higher ($\approx$260) than any other segments, suggesting greater variability in tree topology over the course of MCMC.
SPR distances between replicates of most other segments are $\approx$200.
}
\label{SPRdistancesReplicates}
\end{figure}

\begin{figure}
\centering  
\includegraphics[width=0.65\textwidth]  {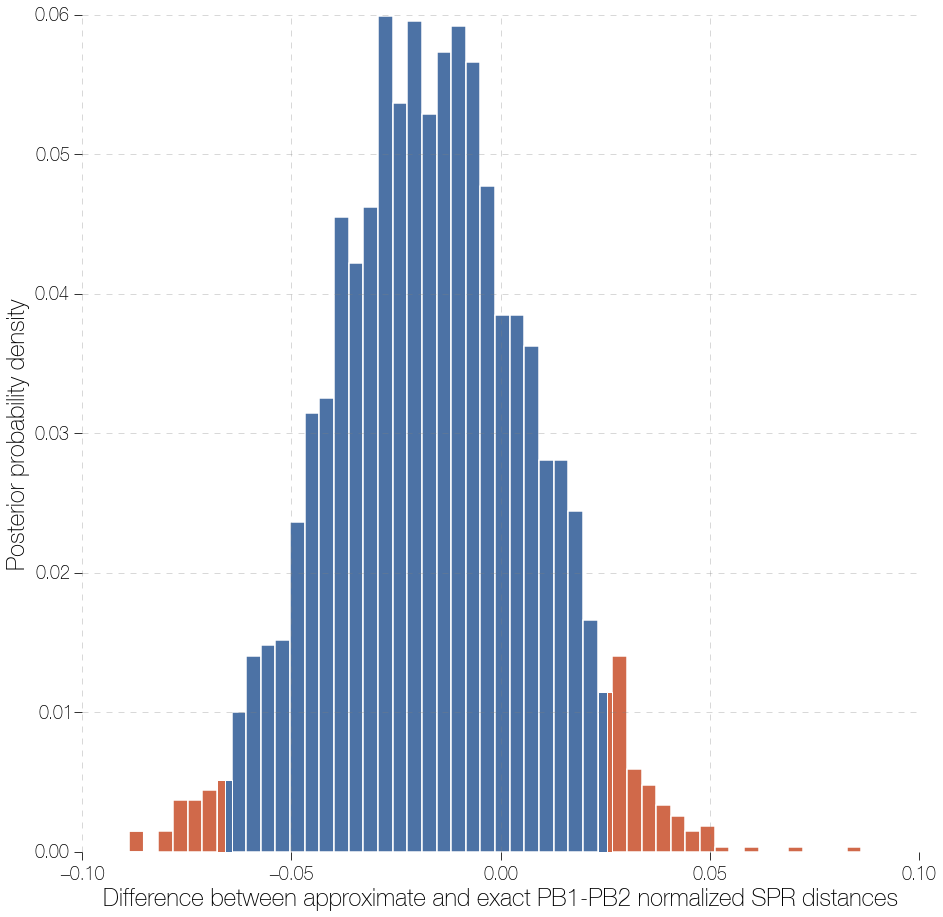}
\caption{\textbf{Distribution of differences between exact and approximate PB1-PB2 SPR distances after normalization.}
95\% HPD interval (blue) overlaps zero, suggesting no evidence of differences between approximate and exact SPR distances following normalization.}
\label{NormSPR_PB1-PB2_difference}
\end{figure}

\begin{figure}
\centering  
\includegraphics[width=0.65\textwidth]  {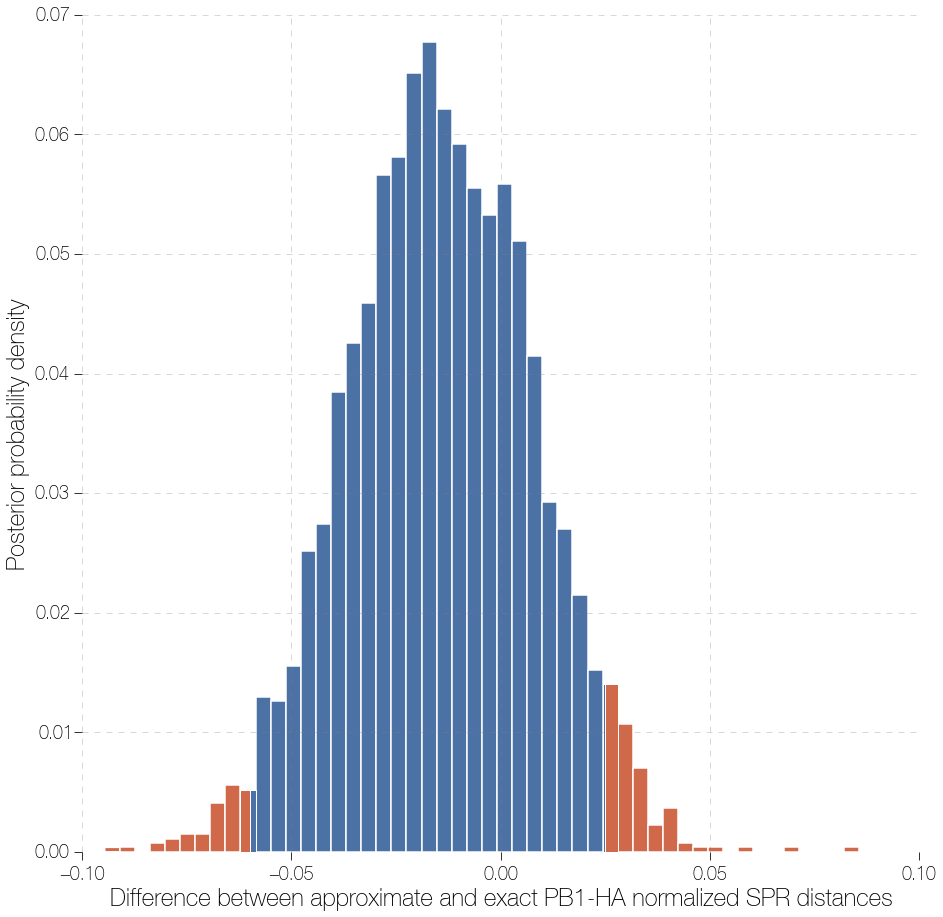}
\caption{\textbf{Distribution of differences between exact and approximate PB1-HA SPR distances after normalization.}
95\% HPD interval (blue) overlaps zero, suggesting no evidence of differences between approximate and exact SPR distances following normalization.}
\label{NormSPR_PB1-HA_difference}
\end{figure}

\begin{figure}
\centering  
\includegraphics[width=0.65\textwidth]  {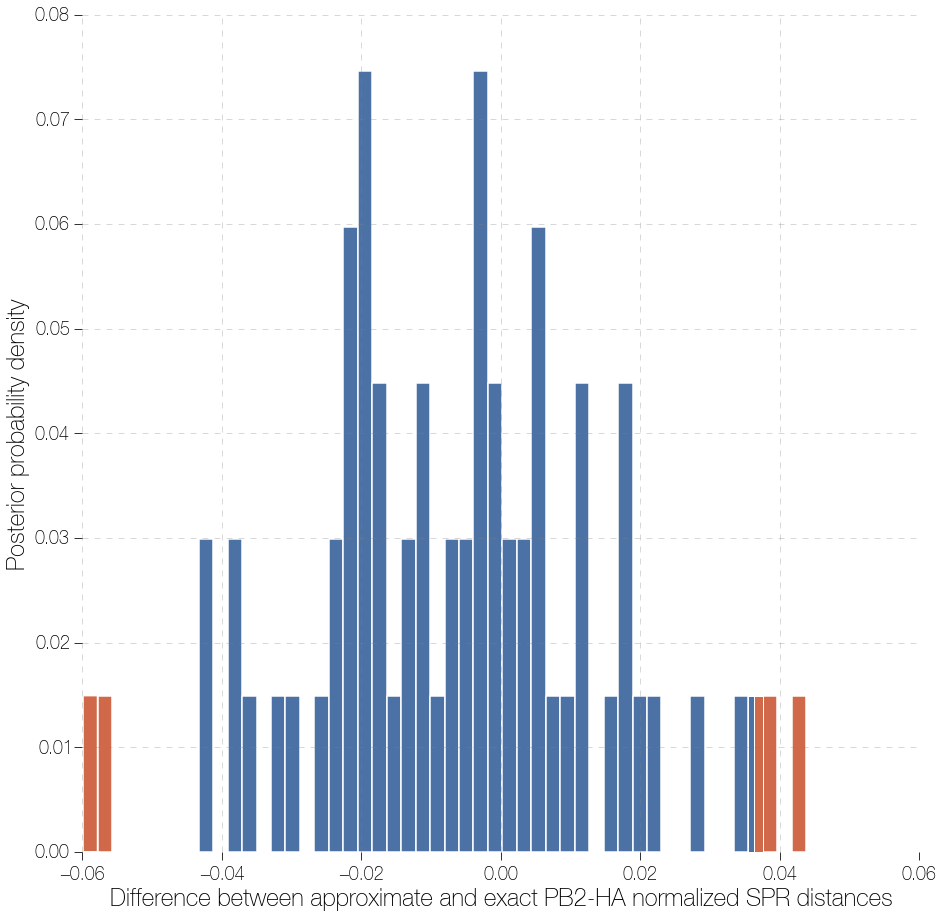}
\caption{\textbf{Distribution of differences between exact and approximate PB2-HA SPR distances after normalization.}
95\% HPD interval (blue) overlaps zero, suggesting no evidence of differences between approximate and exact SPR distances following normalization.
Due to excessively long computation time of exact SPR distances between PB2 and HA trees few comparisons were made.}
\label{NormSPR_PB2-HA_difference}
\end{figure}

The 95\% highest posterior density (HPD) intervals of normalized approximate SPR distances between pairs of segments encompass most means and occupy a relatively small range, suggesting there is no evidence of differences in the number of reassortments between segments (Figure \ref{SPRdistances}).
Reassortment rate given as number of SPR moves per total time in both trees shows similar results (Figure \ref{NormSPR_RErate}).
This is in line with recent experiments in influenza A that have shown that reassortment between segments differing by a single synonymous difference is highly efficient \citep{marshall2013}.
We note, however, that because of phylogenetic uncertainty our estimate of SPR distance might simply lack power.
Comparisons between independent analyses of the same segments yield distances that are comparable to distances between different segments (Figures \ref{SPRdistancesTrees} and \ref{SPRdistancesReplicates}), suggesting that phylogenetic uncertainty is making a considerable contribution to our estimates of approximate SPR distances.
Still, we find that independent replicates from the same segment (Figure \ref{SPRdistancesReplicates}) show lower SPR distances that comparisons between segments (Figure \ref{SPRdistancesTrees}), suggesting that phylogenetic noise is not completely overwhelming reassortment signal.
In addition, SPR distances themselves can only approximate (and underestimate) the actual numbers of reassortments.
Thus we caution against over-interpreting Figure \ref{SPRdistances}.
Although there might be concern about using approximate, rather than exact, SPR distances we do estimate exact SPR distances for a limited number of segment pairs - PB1, PB2 and HA - and find that after normalization exact and approximate SPR distances are not significantly different (Figures \ref{NormSPR_PB1-PB2_difference}--\ref{NormSPR_PB2-HA_difference}).

\begin{figure}
\centering  
\includegraphics[width=0.45\textwidth]  {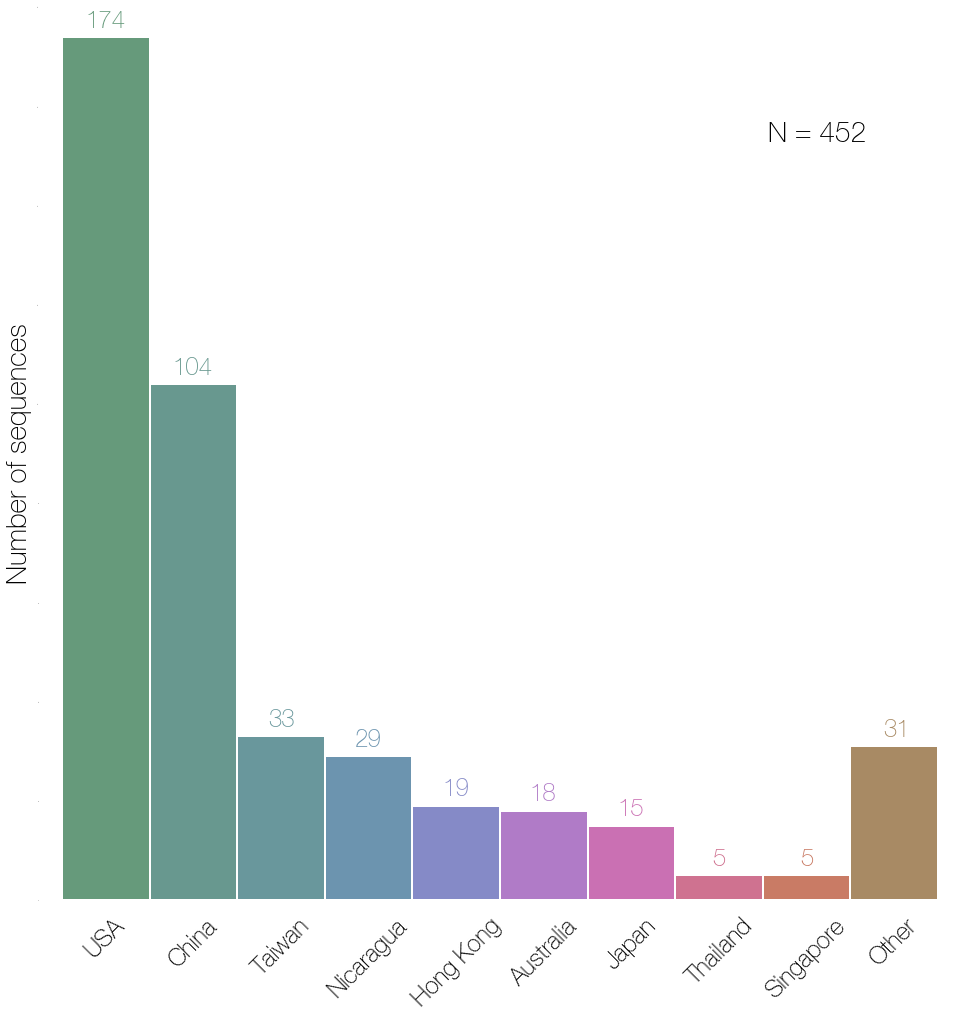}
\caption{\textbf{Geographic distribution of sequences in the primary dataset.}
Sequences were assigned to the ``other'' category if there were less than 5 sequences from that country.
Most of the genomes in the primary dataset were sampled in the USA.}
\label{geoSeqs4}
\end{figure}

\begin{figure}
\centering  
\includegraphics[width=0.45\textwidth]  {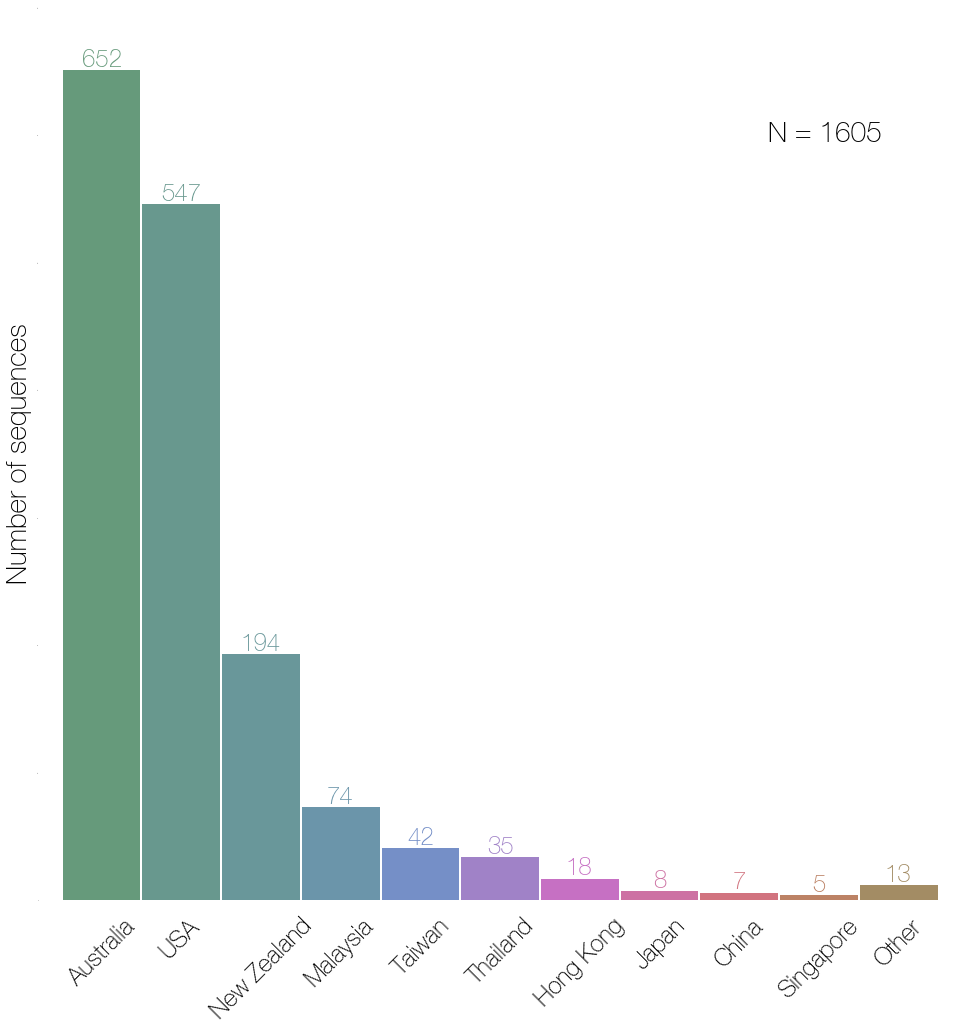}
\caption{\textbf{Geographic distribution of sequences in the secondary dataset.}
Sequences were assigned to the ``other'' category if there were less than 5 sequences from that country.
Most of the genomes in the secondary dataset were sampled in Australia.}
\label{geoSeqs1600}
\end{figure}

\begin{figure}
\centering  
\includegraphics[width=0.75\textwidth]  {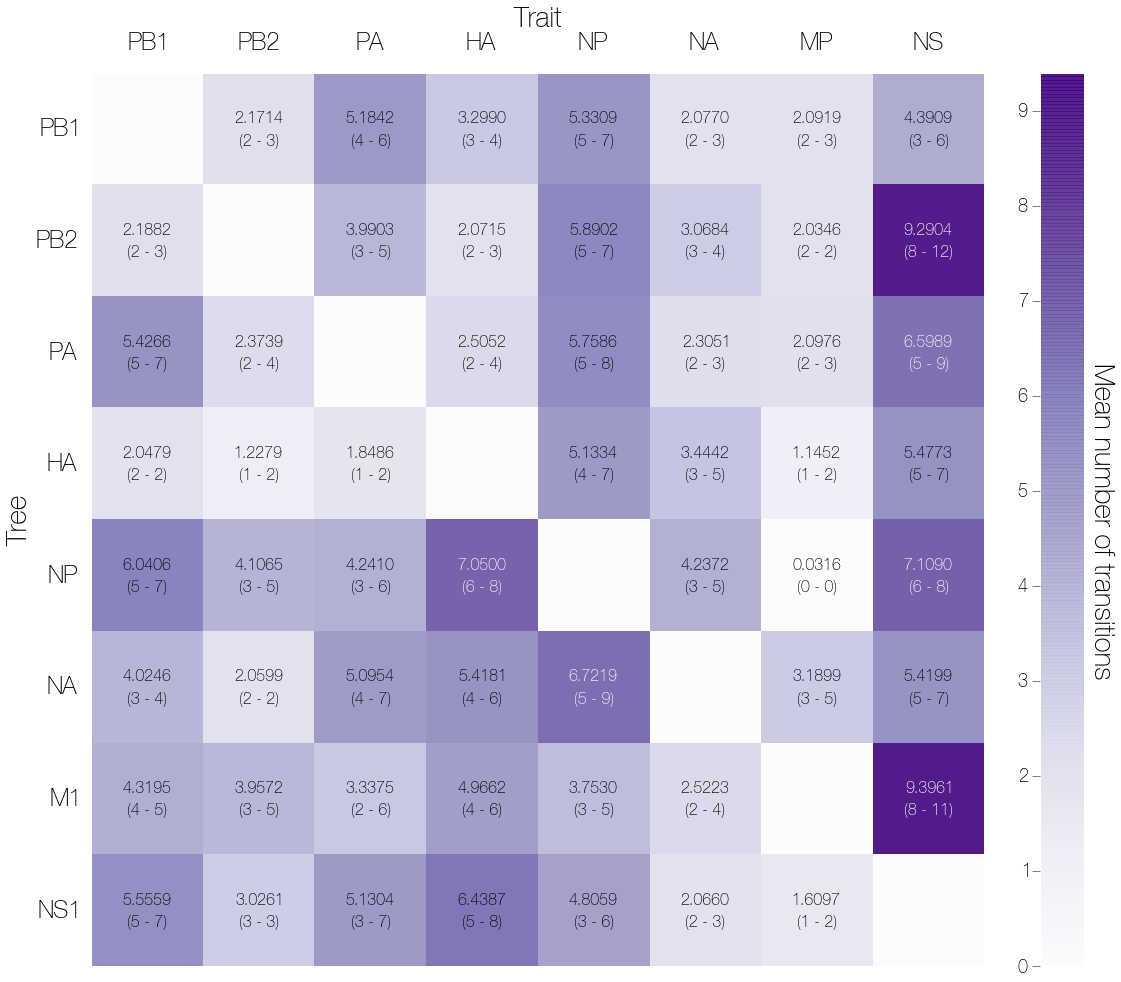}
\caption{\textbf{Numbers of trait transitions in each tree.}
The numbers shown are the mean inferred number of trait transitions (minus one to account for the initial Vic--Yam split) in a given tree and trait combination.
The numbers in brackets correspond to 95\% highest posterior density intervals.
Transitions may not be independent when more than one segment reassorts at the same time.}
\label{Ntransitions}
\end{figure}

\begin{figure}
\centering  
\includegraphics[width=0.65\textwidth]  {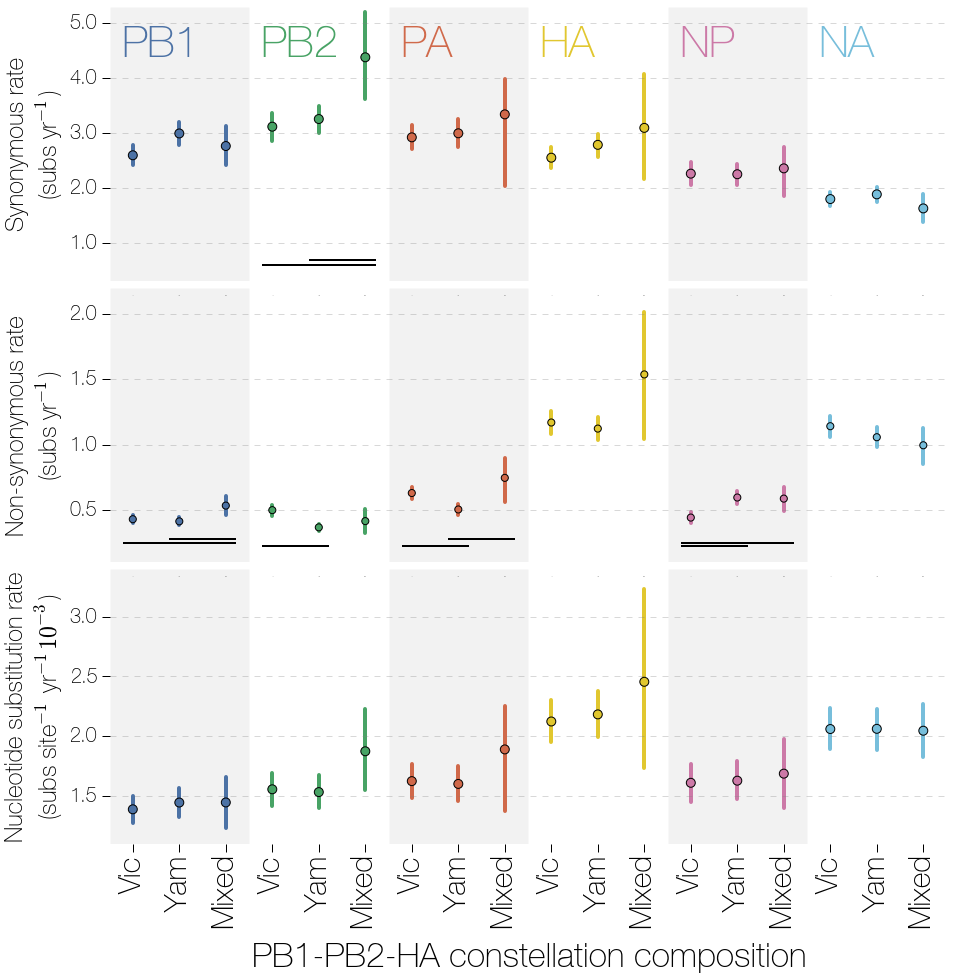}
\caption{\textbf{Synonymous, non-synonymous and nucleotide substitution rates in segments under different PB1-PB2-HA complexes.}
Evolutionary rate dissimilarities under Vic and Yam PB1-PB2-HA complexes are not systematic and appear negligible.
Synonymous and non-synonymous rates were calculated by dividing the sum of all substitutions of a given class by the total amount of evolutionary time under each PB1-PB2-HA constellation.
Nucleotide rates were calculated by multiplying the inferred nucleotide substitution rate on each branch by the branch length, then dividing this by the total amount of evolutionary time under each PB1-PB2-HA constellation.
Vertical bars indicating uncertainty are 95\% HPDs, black bars indicate 95\% HPDs that do not overlap.}
\label{robustCounting}
\end{figure}



\begin{figure}
\centering  
\includegraphics[width=0.65\textwidth]  {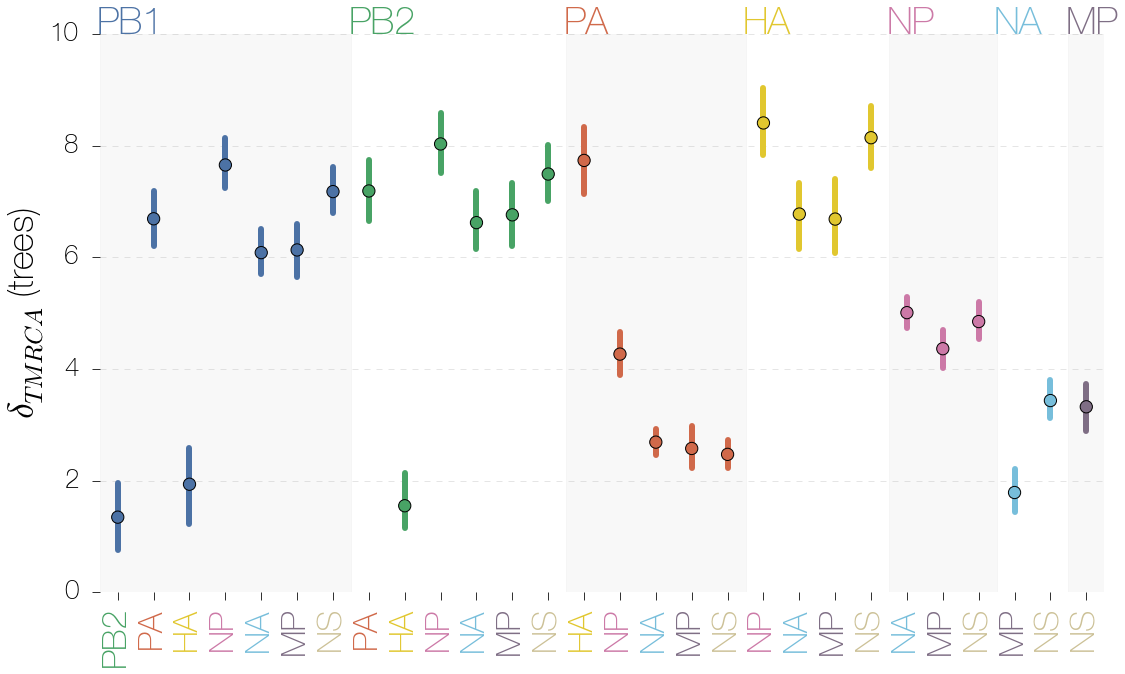}
\caption{\textbf{$\undtmrca$ between all pairs of trees of segments.}
$\undtmrca$ between trees of segments reveal that tip pairs in PB1, PB2 and HA trees have very similar TMRCAs.
The upper tail of the 95\% HPD (HPDs are represented as vertical lines) interval of $\undtmrca$ values for PB1-PB2-HA and MP-NA trees do not exceed 3 years.}
\label{deltaTMRCAtrees}
\end{figure}

\begin{figure}
\centering  
\includegraphics[width=0.65\textwidth]  {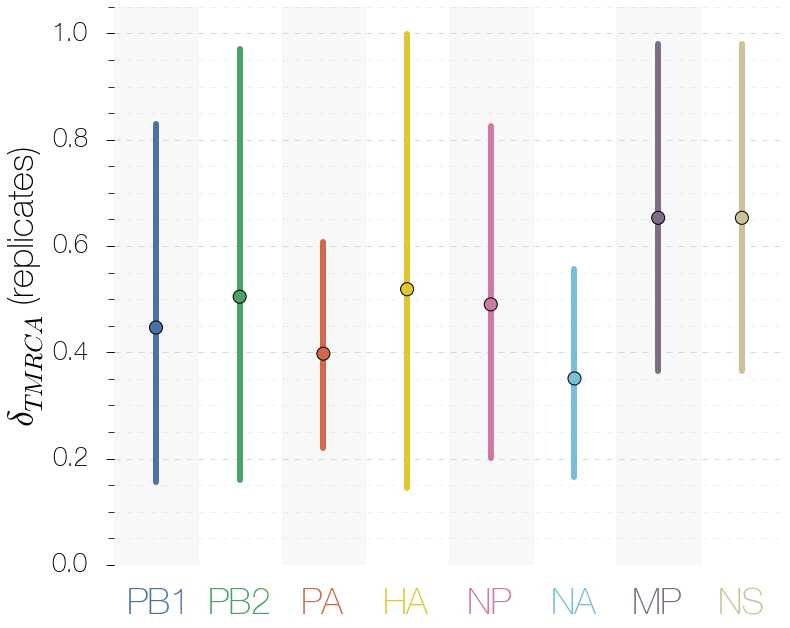}
\caption{\textbf{$\undtmrca$ between replicate trees of each segment.}
$\undtmrca$ values between independent analyses of each segment show that mean $\undtmrca$ values rarely exceed 1 year.}
\label{deltaTMRCAreplicates}
\end{figure}

\begin{figure}[h]
	\centering	
	\includegraphics[width=0.85\textwidth]{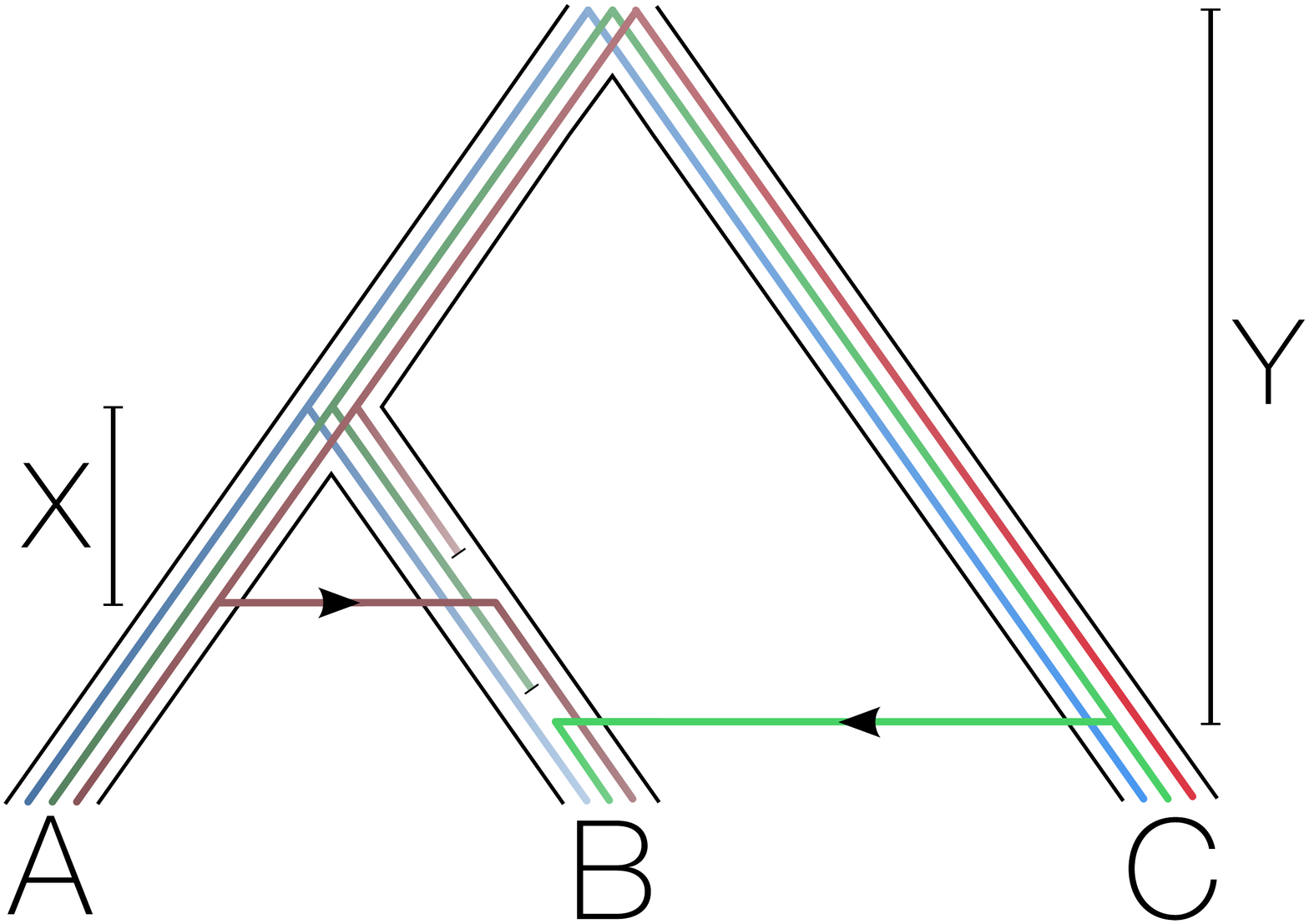}
	\caption{\textbf{Calculating $\undtmrca$ from a species tree perspective.}
Consider an organism that has diverged into 3 taxa (A, B, C) with a genome comprised of 3 segments (blue, green and red).
Due to reassortments taxa A and B share a slightly more recent TMRCA in the red segment, likewise for taxa B and C in the green segment.
By comparing differences in TMRCAs between taxa A-B, A-C and B-C in blue, red and green segments we would find that the red segment has a lower `reassortment distance' (X) than the green segment (Y).
In the absence of reassortment we expect every segment in the genome to have the same tree, \textit{i.e.} the tree of every segment should recapitulate the `virus' tree (analogous to `species' trees in diploid population genetics), including the dates of nodes.
Due to population bottlenecks influenza viruses go through each year we expect strains isolated at any given time to have descended from a single recent virus genome.
This descent from a single genome should therefore be reflected in the TMRCAs of all segments, the only exception being reassortment, which will dramatically alter the TMRCAs of the reassorted segment tree with respect to the background onto which it is reassorting.
}
	\label{speciesTree}
\end{figure}

\clearpage

\bibliographystyle{mbe}
\bibliography{fluB}


\end{document}